\documentclass[%
 reprint,
 amsmath,amssymb,
 aps,
]{revtex4-2}

\usepackage{graphicx}
\usepackage{dcolumn}
\usepackage{bm}
\usepackage{lipsum}
\usepackage{mathrsfs}

\begin{document}

\preprint{APS/123-QED}

\title{A statistical physics and dynamical systems perspective on geophysical extreme events}

\author{D. Faranda}
\altaffiliation{These authors have equally contributed to the manuscript}
\affiliation{Laboratoire des Sciences du Climat et de l'Environnement,  UMR 8212 CEA-CNRS-UVSQ, Universit\'e Paris-Saclay \& IPSL, CEA Saclay l'Orme des Merisiers, 91191, Gif-sur-Yvette, France}
\affiliation{London Mathematical Laboratory, 8 Margravine Gardens, London, W6 8RH, UK}
\affiliation{Laboratoire de Météorologie Dynamique/IPSL, École Normale Supérieure, PSL Research University, Sorbonne Université, École Polytechnique, IP Paris, CNRS, Paris, France} 
\author{G. Messori*}%
\affiliation{Department of Earth Sciences, Uppsala University, Uppsala, Sweden}%
\affiliation{Swedish Centre for Impacts of Climate Extremes (climes), Uppsala University, Uppsala, Sweden}%

\affiliation{Department of Meteorology and Bolin Centre for Climate Research, Stockholm University, Stockholm, Sweden.}
\author{T. Alberti}
\affiliation{Istituto Nazionale di Geofisica e Vulcanologia, via di Vigna Murata 605, 00143, Rome, Italy}%
\author{C. Alvarez-Castro}
\affiliation{Universidad Pablo de Olavide, Seville, Spain}
\affiliation{Fondazione Centro Euro-Mediterraneo sui Cambiamenti Climatici, Bologna, Italy}%
\author{T. Caby}
\affiliation{CMUP, Departamento de Matem\`atica, Faculdade de Ciências, Universidade do Porto,
Rua do Campo Alegre s/n, 4169007 Porto, Portugal.}
\author{L. Cavicchia}
\affiliation{Fondazione Centro Euro-Mediterraneo sui Cambiamenti Climatici, Bologna, Italy}%
\author{E. Coppola}
\affiliation{The Abdus Salam International Center for Theoretical Physics, Trieste, Italy}
\author{R. Donner}
\affiliation{Department of Water, Environment, Construction and Safety, Magdeburg-Stendal University of Applied Sciences, Breitscheidstraße 2, 39114 Magdeburg, Germany}
\affiliation{Potsdam Institute for Climate Impact Research (PIK)—Member of the Leibniz Association, Telegrafenberg A31, 14473 Potsdam, Germany}
\author{B. Dubrulle}
\affiliation{Université Paris-Saclay, CEA, CNRS, SPEC, CEA Saclay 91191 Gif-sur-Yvette CEDEX, France}
\author{V. M. Galfi}
\affiliation{Institute for Environmental Studies, Vrije Universiteit Amsterdam, the Netherlands}
\author{E. Holmberg}
\affiliation{Department of Earth Sciences, Uppsala University, Uppsala, Sweden}
\author{V. Lembo}
\affiliation{Istituto di Scienze dell'Atmosfera e del Clima, Consiglio Nazionale delle Ricerche (CNR-ISAC), Bologna, Italy}
\author{R. Noyelle}
\affiliation{Laboratoire des Sciences du Climat et de l'Environnement,  UMR 8212 CEA-CNRS-UVSQ, Universit\'e Paris-Saclay \& IPSL, CE Saclay l'Orme des Merisiers, 91191, Gif-sur-Yvette, France}
\author{B. Spagnolo}
\affiliation{Dipartimento di Fisica e Chimica ``E. Segr\`e", Group of Interdisciplinary Theoretical Physics, Universit\`a degli Studi di Palermo, I-90128 Palermo, Italy}
\author{D. Valenti}
\affiliation{Dipartimento di Fisica e Chimica ``E. Segr\`e", Group of Interdisciplinary Theoretical Physics, Universit\`a degli Studi di Palermo, I-90128 Palermo, Italy}
\author{S. Vaienti}
\affiliation{Aix Marseille Universit\'e, Universit\'e de Toulon, CNRS, CPT, 13009 Marseille, France} 
\author{C. Wormell}
\affiliation{Mathematical Sciences Institute, The Australian National University, Canberra ACT 2601, Australia}
\author{P. Yiou}
\affiliation{Laboratoire des Sciences du Climat et de l'Environnement,  UMR 8212 CEA-CNRS-UVSQ, Universit\'e Paris-Saclay \& IPSL, CE Saclay l'Orme des Merisiers, 91191, Gif-sur-Yvette, France}

\date{\today}

\begin{abstract}
Statistical physics and dynamical systems theory are key tools to study high-impact geophysical events such as temperature extremes, cyclones, thunderstorms, geomagnetic storms and many more. Despite the intrinsic differences between these events, they all originate as temporary deviations from the typical trajectories of a geophysical system, resulting in well-organised, coherent structures at characteristic spatial and temporal scales. While statistical extreme value analysis techniques are capable to provide return times and probabilities of occurrence of certain geophysical events, they are not apt to account for their underlying physics. Their focus is to compute the probability of occurrence of events that are large or small with respect to some specific observable (e.g. temperature, precipitation, solar wind), rather than to relate rare or extreme phenomena to the underlying anomalous geophysical regimes. This paper outlines this knowledge gap, presenting some related challenges, new formalisms and briefly commenting on how stochastic approaches tailored to the study of extreme geophysical events can help to advance their understanding.

\end{abstract}

\maketitle


\section{Open challenges in the study of geophysical extreme events}

The aim of this perspective is to bridge statistical physics, statistics, dynamical systems theory and geophysics to provide an overview of current techniques suitable to study high-impact events in the earth system. Examples include temperature extremes, cyclones, thunderstorms or geomagnetic storms, all of which can be interpreted as rare states of the underlying geophysical dynamical systems.

The  properties of geophysical extreme events have been extensively studied by using parametric extreme value theory (EVT) approaches \citep[e.g.][]{gumbel_return_1941,davison_models_1990,coles2001,mares_extreme_2009, elvidge_using_2018,french_quantifying_2019,do_nascimento_regression_2021}. Such approaches identify and characterise extremes by identifying their underlying distribution, from which one may for example infer the probability of occurrence of events that are large or small relative to some specific observable (e.g. temperature, precipitation, solar wind).

Nonetheless, conventional EVT techniques come with some limitations (see, e.g.~\cite{HaanFerreira2010}). They are for example not suitable for describing spatially heterogeneous phenomena or phenomena issuing from largely unprecedented dynamics, such as may be the case for some geophysical extreme events. Similarly, conventional EVT does not account for the physical drivers underlying the extremes. This has motivated the introduction of a number of new mathematical formalisms based on defining extreme events as rare recurrences in the phase space of high-dimensional systems~\cite{lucarini2016extremes,ragone2018computation,faranda2018extreme,bouchet2019rare,ghil2020physics,abadi2020dynamical,freitas2020rare,caby2020extreme,haydn2020limiting,pons2020sampling,galfi2021fingerprinting,alberti2021small}. We specifically highlight the work by Lucarini et al. \cite{lucarini2016extremes}, which presents a rigorous framework for the link between extreme value theory and dynamical systems, and \cite{ghil2020physics} which brings together the dynamical systems theory and nonequilibrium statistical physics perspectives. Among the main achievements of the above approaches we note: demonstrating how rare geophysical events correspond to unstable fixed points of the attractor ~\cite{lucarini2016extremes,abadi2020dynamical,freitas2020rare}, identifying special sets of trajectories corresponding to rare events~\cite{ragone2018computation,bouchet2019rare, galfi2021fingerprinting}, detecting partial synchronization of system variables~\cite{faranda2018extreme,haydn2020limiting,pons2020sampling} and elucidating the spatio-temporal variability of dynamical properties of complex systems \cite{caby2020extreme, alberti2021small}. These approaches have recently provided information on a number of geophysical phenomena, including climate extremes and their atmospheric drivers ~\cite{faranda2017dynamical,messori2017dynamical,faranda2019hammam,de2020compound,ragone2021rare, galfi2022persistent,hochman2023sources}, ocean dynamics~\cite{giamalaki2018signatures,giamalaki2021future}, ecosystems~\cite{Grimaudo2022a,Lazzari2021a}, paleoclimate~\cite{messori2021characterising,brunetti2019co}, solar wind turbulence, space weather and exoplanetary atmospheres~\cite{alberti2021small,alberti2022concurrent, hochman2023analogous}, turbulence~\cite{dubrulle2022many} and earthquake dynamics~\cite{gualandi2020predictable,gualandi2022stochastic}.

Here, we seek to provide a detailed perspective of the above techniques and their links to established concepts from dynamical systems theory, which are applied to a range of data from idealised systems to experimental or observational datasets. Examples include the Lorenz-63 attractor~\cite{lorenz1963deterministic}, state-of-the-art climate data such as the ERA5 reanalysis ~\cite{hersbach2020era5} or global climate models~\cite{taylor2012overview,eyring2016overview}, geomagnetic data~\cite{jonkers2003four} and turbulence data~\cite{cappanera2021turbulence}. Considering such a wide range of datasets illustrates how approaches developed for low- and medium-dimensional systems can provide insights into high-dimensional chaotic data. The overview also considers the importance of finite size effects (including space and time resolution of the datasets). We thus bridge the gap between theoretical tools often defined with asymptotic limit theorems and complex datasets. 

While we do not perform a holistic review of all available mathematical and statistical tools for geophysics and geophysical extremes, we briefly outline here some relevant review studies. For example, Lucarini et al~\cite{lucarini2014mathematical} emphasizes the need for a comprehensive understanding of the structural and multiscale properties of climate dynamics. It explores various mathematical and theoretical approaches, including the Nambu formulation of fluid dynamics and statistical mechanics, to construct numerical models and study phenomena like geophysical turbulence and the nonequilibrium nature of the climate system. Another promising theoretical framework for studying geophysical extremes is proposed by \cite{galfi2021applications}. The authors apply Large Deviation Theory, which issues from statistical physics, to problems in geophysical fluid dynamics and climate science.  A broader overview of the physics of climate science, including connections to astrophysics is provided by Ghil and Lucarini\cite{ghil2020physics} who emphasize the complexity and nonequilibrium nature of the climate system, highlighting the importance of natural variability, external forcings, and advanced modeling techniques. 
In the present perspective, we primarily focus on the application of dynamical systems theory and nonequilibrium statistical physics to understand geophysical extremes. As is the nature of a perspective, we highlight selected geophysical extreme events and the associated relevant formalisms.

Our perspective is structured as follows:
\begin{itemize}
    \item In Section 2 we introduce a framework for the study of rare events in dynamical systems, the concept of rarity of a recurrence and its link to the asymptotic theory of extreme events. We further discuss the theoretical problems arising in finite datasets and introduce local dynamical systems indicators for rare events.
    \item In Section 3 we  comment on the stochastic approach useful for investigating extreme geophysical events and explore how stochastic dynamics and phenomena induced by environmental noise can help explain rare geophysical events.
    \item In Section 4 we apply dynamical indicators to understand, model and simulate a variety of geophysical rare events.
    \item In Section 5 we present opportunities and open challenges for the study of geophysical extreme events. We specifically outline lines of research  enabled by recent findings, but that are yet to be explored in the literature.
    \item In Section 6 we discuss practical and theoretical limitations in applying the frameworks described and implemented in the previous sections to geophysical systems.
\end{itemize}  

The paper ends with a summary of key perspectives and findings (Sections 7 and 8).

We conclude this introductory section by providing concise definitions of some of the technical terms used in the later sections of the paper.

\begin{itemize}
\item{Atomic}: A measure $\mu$ (see "Measure" below) is said to be atomic if it gives positive mass to a measurable set, say $A$ (the atom), and all the measurable subsets of $A$ have measure zero. In other words, the measure is concentrated in a small subset of phase-space.
\item{Axiom A}: an Axiom-A diffeomorphism is a smooth map $T$ of a manifold into itself which displays two properties: (i) the invariant set of $T$ (the attractor) is the closure of the periodic points of $T;$  (ii) on the attractor the derivative of $T$ has a uniform splitting into contracting and expanding directions. Crucially, the trajectories of a perturbed Axiom A system have a one-to-one topological correspondence with those of the unperturbed system.
\item{Invariant Measure}: $\mu$ is said to be a $T$-invariant measure in a phase-space $M$ if for all $A\subset M,$ $$\mu(T^{-1}A)=\mu(A).$$ In the dual space of summable functions $f,$ the invariance reads:
$$
\int f(Tx)d\mu(x)=\int f(x)d\mu(x).
$$
\item{Image Measure}: The image measure of the measure $\mu$ by the function $f:M\to \mathbb{R}^k$ is the measure such that for all $A \subset \mathbb{R}^k$, $$\mu_f(A)=\mu(f^{-1}(A)).$$
\item{Information Dimension}: We call the information dimension of the measure $\mu$ the quantity
$$D_1=\lim_{r\to 0}\frac{\int_M \log \mu(B(x,r))d\mu(x)}{\log r},$$ provided the limit exists and 
where $B(x,r)$ denotes a ball of radius $r$ centered at point $x\in M$.
\item{Local dimension}: 
We call the local dimension of the measure $\mu$ at the point $x\in M$ the following quantity:
\begin{equation}
d_\mu(x):=\lim_{r \to 0} \frac{\log\mu(B(x,r))}{\log r},
\end{equation}
provided the limit exists.
\item{Manifold}:
A manifold of dimension $n$  is a space for which  each point has a neighborhood which is the image of  an open subset of 
$\mathbb{R}^n.$
\item{Measure}:
A measure is a function associating a non-negative number to $A\subset M$. Measures additionally need to be countable additive. They are in practice a mathematical generalisation of concepts such as length, area, mass etc.
\item{Penultimate attractor}: The union of the observed (measured or simulated) finite-size trajectories of the system.
\item{Submanifold}: A subset of a manifold which has in turn the structure of a manifold, see above.
\end{itemize}

\section{How to define and track rare events in geophysical datasets}
\label{sect:Define_track}
Mathematically robust limit theorems allow the characterisation of extreme events in idealised systems. Examples include the use of Extreme Value Theory (EVT), including Generalised Extreme Value (GEV) and Generalised Pareto distributions (GPD), to fit block maxima or peaks over threshold. This assumes an infinite timeseries of i.i.d.\ random variables. Extremes arising in real-world dynamical systems however, cannot in principle satisfy these requirements (see Sect. VI). When studying these systems it is thus important to go beyond a pure statistical analysis and also include information on the dynamics. This leads to a distinction between extreme events as defined in a probabilistic way via EVT, and extreme events in a dynamical systems theory context, which can be equated to rare events. In the latter sense, the two terms will hereafter be used interchangeably. 

In order to quantify the rarity of an event, a key tool in dynamical systems theory is the Poincaré recurrence theorem. This theorem states that for a conservative dynamical system, if the system evolves for a sufficiently long time, it will return arbitrarily close to any initial position. In other words, the system will exhibit recurrent behavior and, although rare, certain states of the system will recur infinitely often. Moreover, for $\mu(X, T)$ a measure $\mu$-preserving dynamical system, where $X$ is the phase space and $T$ is the time evolution map, the Poincaré recurrence theorem states that for almost every point $x$ in $X$ and any neighborhood $U$ of $x$, there exists a positive integer $n$ such that $T^n(x)$ (the point $x$ after $n$ iterations of the time evolution map) belongs to $U$. Although the theorem was formulated for conservative dynamical systems, it was later shown that similar conclusions hold for chaotic forced dissipative systems having a strange attractor (e.g. \cite{ruelle1995turbulence}).

The Poincaré recurrence theorem has applications in various fields, including statistical mechanics and ergodic theory. The time required for a trajectory to return to its initial state, known as the recurrence time, is related to the system's phase space volume and the speed of evolution along the trajectory, and can be extremely long for rare states. By studying the recurrence times of trajectories near these rare points, it is possible to gain insights into the underlying dynamics of the system and to identify the most likely pathways leading to the rare events.

The rest of this section provides technical background for the study of rare events in dynamical systems. At the beginning of each subsection, we present a qualitative summary of the key points it touches upon. 

\subsection{Probabilistic theories for extreme events in geophysical time series}

The two main probabilistic approaches to analyse geophysical extreme events are EVT and Large Deviation Theory (LDT). In this section we give a short overview of the basic mathematical formulation for both theoretical frameworks. We also point out practical aspects when applying them to geophysical data.

While EVT deals with tails of probability distributions, LDT considers, in its basic formulation, probabilities of sample averages. From a mathematical perspective, these theories are formulated for i.i.d.\ random variables. However, from a practical perspective, the main requirement is that the extreme values or the sample averages are uncorrelated and homogeneous. This is a reasonable assumption if the system is chaotic enough (see Sect. 3.2 in \cite{lucarini2016extremes}), correlations decay rapidly, and trivial non-stationarities \cite{freitas2017extreme,atnip2022perturbation}, such as seasonal or other regular cycles or anthropogenic climate change signals, are accounted for.

EVT contemplates two main approaches, the \textit{block maxima} and the \textit{peak over threshold} approaches. Both are asymptotic theories of probability distributions of extreme values, with the limit being approached by selecting more and more extreme states. However, the two approaches differ in the way extreme values are selected. 

In case of the \textit{block maxima} approach, extreme values are defined as maxima $M_n=\max\{X_1, X_2, ...\ X_n\}$ of a series of i.i.d.\ random variables $X_1, X_2, ...$ divided into blocks of equal length $n$. If certain conditions are fulfilled, the distribution of properly normalised block maxima $M_n$ converges to a so-called GEV distribution for $n \to \infty$ \cite{coles2001}. In practice, one selects the maxima over fixed time periods, e.g. 1 year, and verifies whether their distribution is properly described by a GEV distribution. The GEV distribution has three parameters: location parameter ($\kappa$), scale parameter ($\sigma$), and shape parameter ($\xi$). The cumulative distribution function of the GEV distribution is given by:

\begin{eqnarray}
G(z) = \exp \left \{-\left [1+\xi \left (  \frac{z-\kappa}{\sigma}\right )\right ]^{-\frac{1}{\xi}} \right \} \hspace{.2cm} \mbox{for}\ \xi \ne 0,\\
G(z) = \exp \left\{  -\exp \left [ -\left (  \frac{z-\kappa}{\sigma}\right )\right] \right\} \hspace{.2cm} \mbox{for}\ \xi=0.
\label{eq:GEV}
\end{eqnarray}
where $1+\frac{\xi (z-\kappa))}{\sigma}>0$, $-\infty <\xi< \infty$, $\sigma>0$ and $-\infty <\kappa< \infty$. 
The sign of the shape parameter determines the type of GEV distribution:
\begin{itemize}
\item $\xi>0$: The distribution has an infinite upper endpoint, and is called the Fréchet distribution. It is used to model distributions with a heavy (power-law) tail decay. 
\item $\xi<0$: The distribution has a finite upper endpoint, and is called the Weibull distribution. It is used to model distributions with a light tail.
\item $\xi=0$: The distribution reduces to the Gumbel distribution, featuring an infinite upper endpoint and exponential tail decay.
\end{itemize}

In case of the \textit{peak over threshold} approach, we consider again i.i.d.\ random variables $X_1, X_2, ...$ and select extreme values as values exceeding a certain high threshold $u$. If $u$ is high enough, the distribution of threshold exceedances $X-u$ follows a GPD \cite{coles2001}. In practice, similarly to the block maxima approach, one verifies that the distribution of exceedances of a high threshold is properly approximated by a GPD. The probability of exceedance of a high threshold $u$ by $X_i$, i.e. $Y_i=X_i-u$ with $Y_i>0$ for $i=1, 2, ...$ is given by a cumulative distribution function, with shape parameter ($\xi$) and scale parameter ($\hat{\sigma}$), that converges to the GPD family:
\begin{eqnarray}
H(y) = 1 - \left ( 1 + \frac{\xi y}{\hat{\sigma}} \right )^{-\frac{1}{\xi}} \hspace{2cm} \mbox{for}\ \xi \ne 0,
\label{eq:GPD0}
\\
H(y) = 1 - \exp \left({-\frac{y}{\hat{\sigma}}}\right) \hspace{2cm} \mbox{for}\ \xi=0,
\label{eq:GPD}
\end{eqnarray}
where $1+\frac{\xi y}{\hat{\sigma}}>0$, $-\infty <\xi< \infty$, $y>0$ and $\hat{\sigma}>0$. In a similar fashion as the GEV distribution, if $\xi=0$ the tail decays exponentially, if $\xi>0$ it decays polynomially and if $\xi<0$ it has an upper bound. The scale parameter refers to the width of the distribution, and is related to the variability of the extremes.

If the GEV or GPD are adequate models of the distribution of extreme values, one can estimate the return periods of very rare and even unobserved extreme events.

LDT deals with the probability of sample averages $A_m=\frac{1}{m}\sum_{i=1}^m X_i$ over blocks of equal length $m$. Under adequate conditions, the probability of $A_m$ decays exponentially for $m \to \infty$:
\begin{equation}\label{eq:LDP}
    \mathbb{P}(A_m=a)\approx e^{-m I(a)},
\end{equation}
where $I(a)\ge 0$ is the so-called rate function \cite{Touchette2009}. If (\ref{eq:LDP}) holds and $I(a)$ has a unique global minimum, the probability of averages decays exponentially everywhere with increasing $m$, except at the mean $\mathbb{E}[A_m]=\bar{a}$ where $I(\bar{a})=0$ and $P(A_m=\bar{a})=1$. This points to the convergence of sample averages to the real mean as described by the law of large numbers \cite{Hollander2000}. In case of applications to geophysical time series, one verifies whether (\ref{eq:LDP}) holds for long enough averaging periods $m \ge m^*$, i.e. whether $I(a_{m>m^*})\approx I(a_{m^*})$. If this is the case, it is possible to estimate the probabilities of averages far away from $\bar{a}$ over time periods longer than $m^*$ based on $I(a_{m^*})$. Hence, one obtains the probability of very rare, even unobserved anomalies from the long-term mean in terms of both their average intensity and duration \cite{galfi2021fingerprinting,galfi2022persistent}. 

EVT primarily focuses on the intensity or magnitude of extreme values. The duration of extreme events can be incorporated in form of the mean cluster size, based on the extremal index (see Eq.~\ref{eq:EI}). LDT considers temporally averaged anomalies, and includes the event duration via the averaging time length. Thus, it is especially useful to study persistent extreme events, such as heatwaves or cold spells, and periods during which extreme values in some observable are unusually frequent \cite{galfi2021fingerprinting,galfi2022persistent}.

The above description of these methods relies on time series of observables and takes a stochastic or probabilistic perspective. A number of geophysical systems, such as the atmosphere or ocean, are (dissipative) chaotic systems. If the system is chaotic enough (see Sect. 3.2 in \cite{lucarini2016extremes}), the stochastic perspective is reasonable and the above methods may be applied. In the following sections, we show how by combining these probabilistic methods with dynamical systems theory, new concepts and methods arise. These have the potential to provide a deeper understanding of the structure of chaotic attractors and the dynamics of geophysical systems.

\subsection{An asymptotic framework to study extreme events in dynamical systems}
\label{sec:IIA}
In this section we elucidate the relation between extreme value theory and the statistical property of recurrence. Given a sufficiently long timeseries, the latter can be investigated on a quantitative basis allowing to derive precise asymptotic behaviours. In particular we explore here two questions, namely: how to determine the typical time needed to observe an extreme event; and the typical duration of such events when they do occur.

In the context of dynamical systems, EVT allows formalizing these two questions in terms of the recurrence statistics of specific system states. Suppose we have a set $U$ in the phase space, of small measure, which we accordingly qualify as a {\em rare set} or as the location of an {\em extreme event.} 
\begin{itemize}
    \item{A first question is: What is the probability that the first visit to $U$  of our physical system is larger than some prescribed time $n$? Suppose now that the system entered the set $U$.}
    \item {A second question is: what is the probability that it resides there $k$ times in a prescribed time interval?}
\end{itemize}
These two questions allow quantifying the rarity of an event and its persistence -- both essential characteristics for understanding geophysical extremes.

To give mathematically workable answers to the above questions, we adopt the framework of discrete dynamical systems. We will consider the temporal evolution given by a discrete dynamical system, or map, $T$. This acts on some compact metric space $M,$ with distance $\text{dist}(\cdot, \cdot),$ and carrying a $T$-invariant measure $\mu$ which will be the underlying (stationary) probability describing the statistical properties of the system. For a dissipative system, this defines the system's attractor. The rare set $U$ will change with $n$ and we will denote it with $U_n,\  n\ge 1$; moreover $\mu(U_n)\rightarrow 0$ when $n$ goes to infinity. In order to get rigorous limit theorems, the sequence $U_n$ is usually taken as monotonically decreasing and converging to a null set $\Lambda$. We now fix a positive number $t$ and we choose the sets $U_n$ in such a way that
 \begin{equation}\label{BL}
 \mu(U_n)=\frac{t}{n}.
   \end{equation}
This is usually possible whenever $\mu$ is not atomic, namely a measure concentrated at a single point, and the sets $U_n $ are sufficiently regular, for instance they are balls shrinking around a point or strips collapsing on a smooth submanifold.  We then define, for $x\in M$:

 $$
 \tau_{U_n}(x):=\inf\{k\ge 1, T^kx\in U_n\},
 $$
which is the first time the initial point $x$ enters the set $U_n.$
To address our first question, we compute:

$$
\mu(\tau_{U_n}>n).
$$

The use of EVT combined with strong chaotic properties of the dynamical systems $(T, \mu)$ allows us to prove the existence of a positive number $\theta\in [0,1]$ such that
\begin{equation}\label{ge}
\lim_{n\rightarrow \infty} \mu(\tau_{U_n}>n)= e^{-\theta  t}.
\end{equation}
This is Gumbel's law. The number $\theta$ is called the {\em extremal index (EI)}. We refer the reader to~\cite{lucarini2016extremes} and  \cite{caby2020extreme} for a detailed discussion of the dynamical systems for which the previous result holds; in a few words: they are hyperbolic dynamical systems with exponential decay of correlations.

The EI can be related to a local persistence indicator \cite{moloney2019overview}, suitable to estimate the average cluster size of the trajectories within the neighborhood of the null set $\Lambda.$ Such an index is in fact less than one when clusters of successive recurrences happen, which is the case, for instance, when the set $U_n$ shrinks around a periodic point or, more generally, around an invariant submanifold. Given a time interval $\Delta,$ the distribution of $\lambda_l$ for a cluster of size $l$, can be defined as the frequency of visiting $l$ times the set $U_n$. We give in the Appendix the analytic expression of $\lambda_l.$ This holds in the limit of $n\rightarrow \infty$ and $\Delta\rightarrow \infty$. It can then be shown that \cite{haydn2020limiting}:

\begin{equation}
\theta^{-1}=\sum_ll\lambda_l.
\label{eq:EI}
\end{equation}

In other words, the EI can be interpreted as the inverse of the expected cluster size of recurrences about $U_n$. For the strongly chaotic systems quoted above, the EI can be computed by estimating suitable return time functions. We in fact have the formula:
\begin{equation}\label{EI}
\theta=1-\sum_{k=0}^{\infty}q_k,
\end{equation}
with $q_k=\lim_{n\rightarrow \infty}q_{k,n}$, provided the limit exist, 
and where the analytic expression of the $q_{k,n}$ is given in the Appendix. This result is based on a perturbative technique introduced by Keller and Liverani \cite{keller2009rare,keller2012rare}. It has been used in the context of EVT for both deterministic and random systems, providing very efficient explicit formulae for the computation of the extremal index \cite{caby2019generalized, caby2020extreme, caby2020computation}. In practice, in the limit of an infinite timeseries, $\theta$ differs from 1 only at periodic points, where $\theta <1$. However, when estimated at finite resolution $\theta$ differs from 1, and its value indicates the stickiness of the state of the system being analysed (see Sect. \ref{sec:IIC}). 

Let us now suppose that the set $U_n$ is a ball $B(z, e^{-u_n})$ of center $z\in M$ and radius $e^{-u_n}$ with $\mu(B(z, e^{-u_n}))=\frac tn$ and $u_n\rightarrow \infty, n\rightarrow \infty.$ If we define the process $X_i(x):= -\log \text{dist}(T^ix, z)$, we can then elucidate the connection between hitting times and EVT as:
\begin{equation}\label{et}
\lim_{n\rightarrow \infty}\mu(M_n\le u_n)=\lim_{n\rightarrow \infty} \mu(\tau_{B(z, e^{-u_n})}>n)= e^{-\theta  t},
\end{equation}
where
$M_n=\max(X_0, X_1,\dots, X_{n-1}).
$ The condition $\mu(B(z, e^{-u_n}))=\frac{t}{n}$ now becomes, for large $n$, 
\begin{equation}\label{ee}
e^{-u_n d_{\mu}(z)}\sim \frac tn,
\end{equation}

where $d_{\mu}(z)$ is an estimate of the local dimension at the point $z.$ Leveraging GEV theory \cite{faranda2011numerical}, it is possible to fit the boundary level $u_n,$ and therefore the dimension $d_{\mu}(z),$ provided that the convergence in (\ref{et}) holds. The EI $\theta$ computed with formula (\ref{EI}) will also be a function of the point $z,$ and we therefore write $\theta(z).$ The couple $(d_{\mu}(z), \theta(z))$ allows us to characterise the local structure of attractors. In~\cite{lucarini2012universal} it was shown that an alternative approach for the selection of the maxima is to use the peak over threshold approach, allowing to estimate $d_{\mu}(z)$ by leveraging the GPD. This enables computing $d$ for a larger class of dynamical systems than those satisfying the conditions for the block maxima approach \cite{lucarini2016extremes}. In particular, the GPD approach is suitable for non-axiom A chaotic dynamical systems with multifractal properties, periodic systems and quasi-periodic systems with the exclusion of periodic points \cite{lucarini2012universal}. The above include a large number of geophysical systems (see the examples provided in the Introduction).\\

We now come to our second question. The orbit of our dynamical system visits infinitely often the set $U_n.$ We could therefore
expect that the exponential law $e^{-\theta t}$ given by the extreme value distribution describes the time between successive
events in a Poisson process. To formalize this, we introduce the random variable
\begin{equation}\label{PPP}
\mathcal{N}^{(n)}(t):=\sum_{j=0}^{\frac{t}{\mu(U_n)}}{\bf 1}_{U_n}\circ T^j,
\end{equation}
and we consider the following distribution:
\begin{equation}\label{PQQ}\mu(\mathcal{N}^{(n)}(t)=k).
\end{equation}

 For a large class of dynamical systems, it has been proven in \cite{haydn2017almost} and also in \cite{haydn2020limiting} that
$$
\mu(\mathcal{N}^{(n)}(t)=k)\rightarrow \nu(\{k\}),
$$
as $n\rightarrow \infty,$ where $\nu$ is the compound Poisson distribution for the parameters $s\lambda_l$, where
$s=\theta t.$
The generating function of a compound Poisson distribution $\nu$ with parameters $s\lambda_k$
reads:
\[
\varphi_{\nu}(z)=e^{\sum_{k\ge1}s\lambda_k\left(z^k-1\right)}.
\]

When $\lambda_1=\lambda$ and $\lambda_k=0$, $k\ge2$, we obtain the Poisson distribution with parameter $t\lambda$. An integer valued random variable $W$ is compound Poisson distributed if there are i.i.d.\ integer valued random variables $\{X_i\}_{i\ge 1},$ and an independent Poisson distributed random variable $N$ so that $W =\sum_{j=1}^NX_j$. The Poisson distribution of $N$ describes the distribution of clusters whose sizes are given by the random variables $X_j$ whose probability densities are the  values $\lambda_l=\mathbb{P}(X_j)=l,\ l=1,2,\dots$. Therefore, for the class of chaotic systems considered here, we get the distribution of clusters which occur on a large  timescale, and the number of returns in each cluster is given by the random variables $X_j$. These returns are on a fixed timescale and nearly independent of the size of the return set as its measure is shrunk to zero.

For instance, when  $\lambda_\ell=(1-p)p^{\ell-1}$ we have that $\theta=1-p$ and    $N\sim\text{Poisson}(t(1-p))$. This specific case is termed \emph{P\'olya-Aeppli}
distribution, with parameter $t(1-p)$. The mass distribution $\nu(\{k\})$ of such a distribution is recalled in the Appendix.   
It is interesting to note that we have two families of indices to compute respectively the EI and the compound Poisson, namely the $q_k$ and the $\lambda_l.$ The two are related since $\theta^{-1}=(1-\sum_{k=0}^{\infty}q_k)^{-1}=\sum_l l\lambda_l,$  which shows, as we already pointed out, that the EI is the inverse of the expected cluster size. This leads to a whole spectrum of $\lambda_l,$ or alternatively of the associated compound Poisson statistics, providing insights into  the distribution of the clusters even in systems with multiple timescales \cite{gallo2021number, haydn2020limiting, caby2020extreme}. For example, \cite{caby2020extreme} studied the statistics of the number of visits of a given observable in the neighborhood of a particular value, for a vector containing the values of sea-level pressure on a grid of $\approx 100$ locations over the North Atlantic. The resulting distribution was very close to P\`olya–Aeppli (see Fig.~11 in \cite{caby2020extreme}), with an excellent agreement with the observational data.

\subsection{Finite time approaches to study rare events in dynamical systems}
\label{sect:finite}

Many geophysical dataseries are of limited length or focus on specific parts of a more complex system. This hinders the direct application of asymptotic results to this data. In this section, we discuss how the analysis tools discussed in the previous sections may be used for the study of high-dimensional complex systems.
 
Formally, the results of EVT are stated with respect to the physical invariant measure $\mu$ supported on the system's attractor $M$, and are asymptotic results. Nonetheless, the characteristics of the actual attractor of the system can be deduced from those of its penultimate attractor, using the framework of large deviations. We will here state results concerning the local dimensions as a paradigmatic example, but this framework can be applied to other dynamical quantities of interest, such as recurrence times, which are closely related objects \cite{caby2019generalized}. The local dimension at a point $x$ in phase space is defined as 

\begin{equation}
d_\mu(x)=\lim_{r \to 0} \frac{\log\mu(B(x,r))}{\log r},
\end{equation}
\noindent
where $B(x,r)$ as before denotes a ball centered at $x$ of radius $r$. Its pre-asymptotic version 

\begin{equation}\label{fid}
d_{\mu,r}(x)= \frac{\log\mu(B(x,r))}{\log r},
\end{equation}
\noindent
converges for almost all states $x\in M$ to the so-called information dimension $D_1$ of the system as $r\to 0$.  This happens  for ergodic exact-dimensional systems and therefore the local dimensions $d_\mu$    will be  almost everywhere  constant with common value  $D_1.$
Depending on the length of the available time series, one can fix a small radius $r>0$ and evaluate $d_{\mu,r}$ at this finite resolution, for example by fitting the empirical distribution of the maximum associated with a suitable observable. When the radius $r$ is small, but not zero, (large) deviations from the typical value $D_1$ can be observed and  they decay exponentially fast  with a  rate function given by \cite{coutinho2018large}:
\begin{equation}
Q(s)=\sup_{q\in \mathbb{R}}\{-qs+qD_{q+1}\},
\end{equation}
\noindent
 which is convex and vanishes at $D_1$. $Q$ depends only on the generalized dimensions of the system, defined for $q\neq 1$ by:

\begin{equation}
D_q=\lim_{r\to 0}\frac{\log\int_M \mu(B(x,r))^{q-1} d\mu(x)}{(q-1)\log r}.
\end{equation}

We note that \cite{datseris2023estimating} perform a related calculation but computing the distance between different pairs of points rather than selecting a single point $\zeta$, and thereby obtain $D_2$. This is in turn similar to the approach in \cite{faranda2018correlation}.

The spectrum of dimensions $D_{q}$ is non trivial for systems exhibiting a wide variety of scaling behaviors, as expected for geophysical systems such as the climate system. A large deviation relation is  given explicitly in \cite{caby2019generalized} and reads:
\begin{equation}
\mu\left(\left\{z \in M \mbox{ s.t. } d_{\mu}(z,r)\in I\right\}\right)\underset{r\to 0}{\sim} r^{\inf_{s\in I}Q(s)},
\end{equation}
\noindent
where $I$ denotes an open interval around $D_1$. In the sense of the above formula, the wide distribution of (finite-resolution) local dimensions over the phase space originates from the multi-fractal structure of the attractor. We see from (\ref{fid}) that for a given radius $r$, the value of $d_{\mu,r}(x)$ is determined by the measure of a ball centered at $x$. Smaller values of $d_{\mu,r}$ correspond to regions of the attractor that are more visited by the dynamics, whereas higher values correspond to less dense regions of the attractor.

Partial observation of the system can be modeled by an observable $f:M\to \mathbb{R}^k$ that acts as a projection from the phase space of the system to the observational space $\mathbb{R}^k$. Examples of such $f$ include gridded observables, delay-coordinate observables used in embedding techniques, or a concatenation of various scalar observables of interest. The measuring process consists in collecting the values of $f$ along a typical trajectory of the system. The obtained data are sampled with respect to the image measure $\mu_f$. This measure is such that for all $A \subset \mathbb{R}^k$, $$\mu_f(A)=\mu(f^{-1}(A)).$$ The statistical properties of the observable $f$ are governed by the geometric structure of the support of this measure. In particular, the recurrence properties are modulated by the local dimensions, and the synchronization properties (several or all variables of the system attain similar values) are governed by the generalized dimensions of $\mu_f$ \cite{caby2020extreme,caby2022matching}. 

It is crucial to understand the relations between the properties of these observations and the ones of the underlying physical system that is being observed. The following result, by \cite{hunt1997projections}, provides an insight into the matter. For a generic differentiable observable \footnote{This equation holds for a prevalent set of observables (see ~\cite{hunt1997projections}).} $f:M \to\mathbb{R}^k$ and for a generic point $x\in M$, the local dimension of the image measure at $f(x)$ is given by

\begin{equation}\label{hk}
d_{\mu_f}(f(x))=\min(k,d_\mu(x)).
\end{equation}

From here, for a generic observable, there are two possibilities:

\begin{itemize}
\item If $k < d(x)$, then $d_{\mu_f}(f(x))=k$, and the information on the local structure of the underlying attractor is lost.
\vspace{\baselineskip}
\item If $k$ is large enough, the information dimension of the underlying attractor is preserved, since for $\mu$-almost all $x\in M$: $$d_{\mu_f}(f(x))=d_\mu(x)$$.
\end{itemize}

For instance, if we properly decompose a dynamical system into $K$ modes, and estimate for a growing number of $k$ modes the dimensions $d_{\mu_f}$, as $k\to K$  we approach the dimensions of the underlying attractor \cite{alberti2022novel}.\\

\subsection{Local dynamical indicators for rare events}
\label{sec:IIC}

As discussed in Sect. \ref{sect:finite}, applications of dynamical systems theory to geophysical data must account for short or incomplete sets of data. Here, we investigate whether two local dynamical systems indicators and associated second-order statistics can provide information on data-limited chaotic systems.

The estimation of the  local dimension $d$, mentioned in Sect. \ref{sect:finite} above rests on the application of extreme value theory to Poincar\'e recurrences in dynamical systems. This approach was first introduced in \cite{lucarini2016extremes} and \cite{faranda2017dynamical}. For a given point $z$ in phase space (e.g., a given sea-level pressure map over a given geographical domain), one computes the probability that the system returns within a ball of radius $e^{-u}$ centered on the point $z$. To do so we first define negative logarithmic returns as:

\begin{equation}
g(x(t))=-\log(\textrm{dist}(x(t),z)) 
\end{equation}
\noindent

Requiring that a point on the orbit falls within a ball of radius $e^{-u}$ around the point $z$ is equivalent to asking that the corresponding value of the series $g(x(t))$ is above the threshold $u$. Under the hypothesis of independence of the exceedances, we leverage the theorem from \cite{freitas2010hitting} to obtain:


\begin{equation}
\mu((X - u(q)) > y| X\ge u(q)) \approx \exp \left(-\frac{y}{\sigma}\right),
\label{eq:GPD2}
\end{equation}

where $u(q)$ is a high threshold associated to a quantile $q$ of the series $X\equiv g(x(t))$. As observed by \cite{lucarini2012universal} the resulting distribution is the exponential member of the GPD family. The parameter $\sigma$, namely the scale parameter of the distribution, depends on the point $z$ in phase space and, for finite time series, on $t$. The local dimension $d(z)$ can be obtained via the relation $\sigma=1/d(z)$. This is the local dimension introduced in Sect. \ref{sect:finite}. In practice, to compute the local dimension, one explicitly fits an exponential distribution to the exceedances above $u(q)$ (see discussions in \cite{datseris2023estimating} and \cite{pons2023statistical}). If, instead of assuming an exponential distribution, one fits a GPD to the data, then the resulting $\sigma$ will not be the inverse of a local dimension, as it will also be a function of the distribution's shape parameter $\xi$.

By combining eq. \ref{et} with eq. \ref{ee}, we can obtain:

\begin{equation}
\mu(M_n\le y)\approx e^{-\theta  (\log n-\sigma y)},
\end{equation}
\noindent

Where we recall that $M_n=\max(X_0, X_1,\dots, X_{n-1})$. A metric of persistence can then be obtained as the inverse of the extremal index $\theta$, dimensionalised by the timestep of the data being used. As also shown by \cite{freitas2012extremal}, in presence of clustering the extreme value law corresponding to the asymptotic limit of recurrences within a neighbourhood around a given phase-space point is modulated by $\theta$ (see their eq. 2.8).

 The local dimension $d$ and persistence $\theta$ thus provide complementary information. $\theta$ relates to the dynamics around a point, and its calculation depends on preserving the chronological order of the data. The extremal index can be estimated following \cite{suveges2007likelihood}, who proposes a maximum likelihood estimator based on block maxima. $d$ instead gives the scaling of the measure around a point (and its calculation is insensitive to time reshuffling of the data). $d$ can be computed using either block maxima or peaks over threshold, and the two can be connected in the asymptotic limit through the result of \cite[][p. 75]{coles2001}: “[...] if block maxima have approximating distribution G [N.B. here G is the GEV], then threshold excesses have a corresponding approximated distribution within the generalized Pareto family". The POT formulation including $\theta$ in the exponent, first introduced in Eq. 4 by \cite{faranda2017dynamical}, thus only holds in the asymptotic limit. For practical purposes, $\theta$ is not estimated using POT, as explained above.

The above framework can also be applied in a bivariate context. Given two observables $x(t)$ and $y(t)$, one can define the state $z=\{z_x, z_y\}$. The joint negative logarithmic returns can then be expressed as: 

\begin{widetext}
\begin{equation}
g(x(t),y(t))=-\log \left[
\mathrm{dist}\left(\frac{x(t)}{||x||},\frac{z_x}{||x||}\right)^2 + \mathrm{dist}\left(\frac{y(t)}{||y||},\frac{z_y}{||y||}\right)^2
\right]^\frac{1}{2}
\label{eq-bivariateg}
\end{equation}
\end{widetext}

where $||\cdot||$ is the average root mean square norm of the coordinates of a vector. For example, $||x||=E_t\left( \left[\sum_i^K x_i(t)^2\right]^{\frac{1}{2}}\right)$, where $K$ is the number of components of $x$ and $E_t$ is an average over time $t$. Based on Eq. (\ref{eq-bivariateg}), one can then compute the co-dimension $d_{x,y}$ and inverse co-persistence $\theta_{x,y}$.

Finally, in a bivariate setting one can define an additional indicator, which we term co-recurrence ratio, as:

\begin{equation}
\alpha(z)=\frac{\nu[g(x(t))>s_x(q) \cap g(y(t))>s_y(q)]}{\nu[g(x(t))>s_x(q)]}
\end{equation}

with $0\leq \alpha \leq 1$, provided that the same number of recurrences is defined for both observables. Here, $\nu[-]$ is the number of events satisfying condition $[-]$, and all other variables are defined as before. Thus, $\alpha$ quantifies how often the two observables have joint recurrences, namely their co-recurrence. If the same number of recurrences is defined for both observables, then by definition $\alpha$ is symmetric with respect to the choice of variable ($x$ or  $y$), since $\nu[g(x(t))>s_x(q)]\equiv\nu[g(y(t))>s_y(q)]$.

The local dimension, persistence and co-recurrence ratio can be computed at each point in the phase space of a physical system, including when the phase space is imperfectly sampled. In the limit of an infinite number of infinitely long trajectories, the local dimension of all points on the attractor will almost surely tend to the information dimension of the attractor (see Sect. \ref{sect:finite}). With a finite size sample, this is not the case and the value of the local dimension computed at each point gives information on the recurrences of these points in the phase space, namely on the geometric characteristics of the trajectory \cite{lucarini2016extremes}.

The above dynamical systems metrics are typically estimated without taking explicitly into account the continuous nature of the trajectories representing the evolution of geophysical systems (see e.g. \cite{faranda2017dynamical}). We propose here an estimator for the local dimension $d$ targeted to continuous dynamical systems, and investigate its second order properties (i.e. we compute its gradient in the phase space). The new estimator uses the $N$ closest recurrences of $z$ within a radius $e^{-u}$, with $u$ given. The local dimension $d_u(z)$ is given by:

\begin{equation} 
    d_u(z) =\frac{\sum_{n=1}^N v_n^{-1} l_n}{ \sum_{n=1}^N v_n^{-1}( l_n - r_n \cos^{-1} (e^u r_n))},
    \label{eq:ld}
 \end{equation}
where $v_n$ is the speed in the phase space at the $n$-th closest recurrence, $r_n$ is the distance between the point $z$ and the $n$-th closest recurrence and $l_n=\sqrt{e^{-2u} - r_n^2}$.

We emphasise that even assuming infinite data, this is only a consistent estimator as $u\to 0$ for measures with a continuous density on a submanifold, for example those generated by stochastic processes. For fractal attractors it will converge only in the sense of C\'esaro averaging, and even in that setting is not known to be consistent. Nevertheless, at finite resolution it provides useful information about the scaling of the system.

\begin{figure}[t]
  \noindent\includegraphics[width=0.5\textwidth]{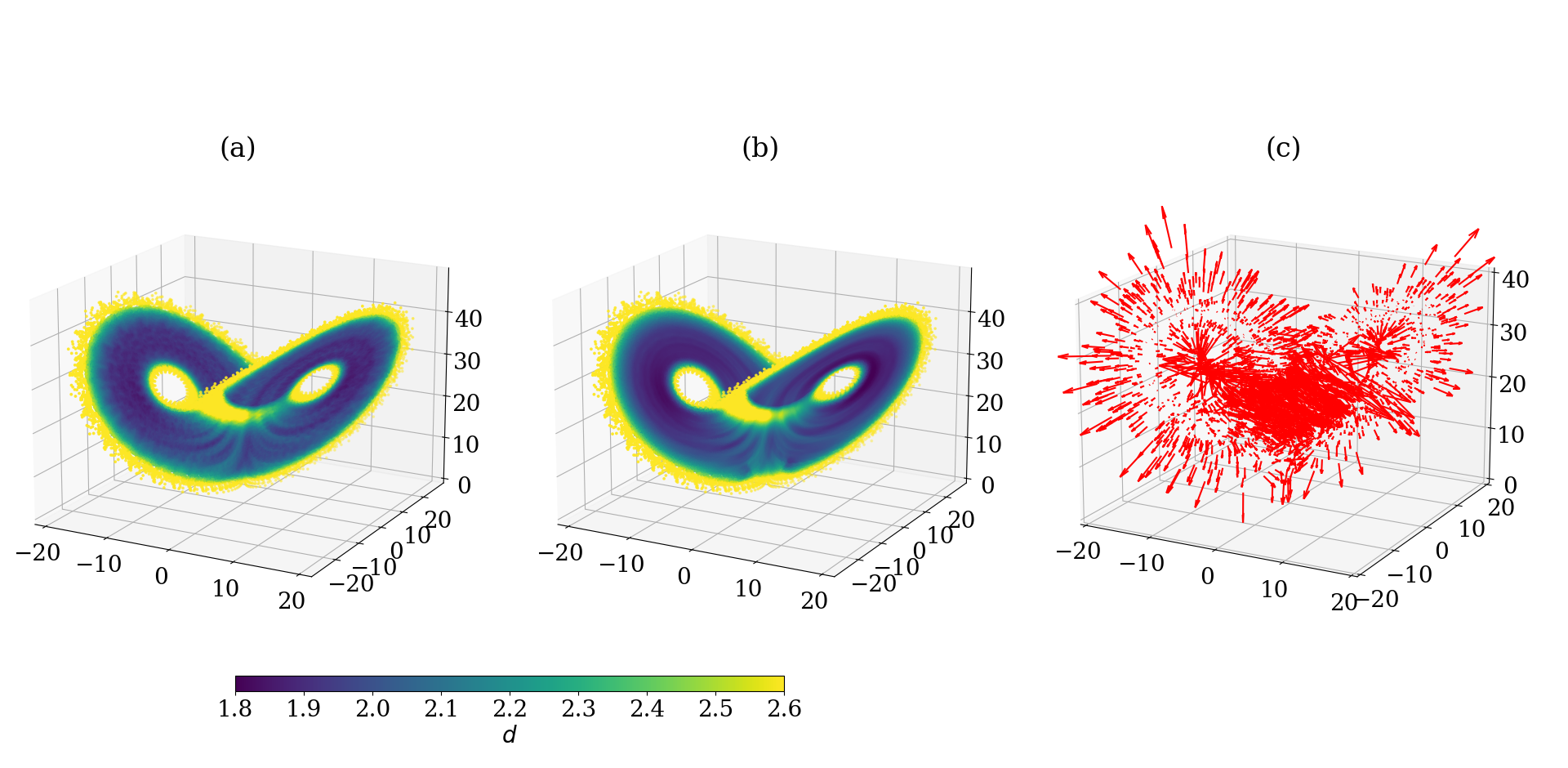}\\
  \caption{\textbf{Gradient of local dimension for the Lorenz-63 system.} (a) Local dimension of the Lorenz-63 model using the GPD estimator; and (b) the estimates based on Eq.~\ref{eq:ld}. (c) Corresponding gradient of the local dimension.} 
  \label{fig:Lorenz63_gradient}
\end{figure}

In a simple low-dimensional attractor such as for the Lorenz-63 system \cite{lorenz1963deterministic}, one can visualize the regions with low and high dimensions (cf.\ Fig.~\ref{fig:Lorenz63_gradient}a, b). The two estimates of local dimension based on the GPD and on Eq.~\ref{eq:ld} are essentially indistinguishable. We further note that for a point situated in the low-dimensional part of one of the wings of the attractor, its local dimension can be increased by moving either towards the exterior of the wing or towards the fixed point in the middle of the wing. 

However, in a high-dimensional system, it is far from easy to know which direction in the phase space would increase or decrease the local dimension, i.e. which are the common and rare structures with respect to the sampling provided by the data set. We therefore propose an estimator of the gradient in the phase space of the quantity $d_u(z)$ defined above.

The gradient $\nabla_{z} d_u(z)$ is a vector in a phase space which has the point $z$ as its origin. It therefore points towards the direction of increasing local dimension relative to the point $z$. We show the directions of the normalized gradients $\frac{\nabla_{z} d_u(z)}{||\nabla_{z} d_u(z)||}$ for the Lorenz '63 model in Fig.~\ref{fig:Lorenz63_gradient}c. The vectors at each phase space point match the intuition with regards to the distribution of local dimensions in Fig.~\ref{fig:Lorenz63_gradient}b.

When computed on a data set with finite size, i.e. undersampling the true attractor of the dynamical system studied, the local dimension thus provides information on the clustering of nearest neighbors in the phase space. Here we proposed to expand this characterization by computing a second order statistics, namely the gradient in phase space of local dimension. We illustrate how this can inform on high-dimensional geophysical systems in Sect. \ref{sect:rare_gradient}.

\section{Stochastic dynamics for understanding and simulating geophysical rare events}

The modeling of chaotic systems is a challenging task because these systems are highly sensitive to initial conditions and external forcings, and small perturbations can lead to significant changes in their behavior. Examples include modulation of the fine structure of the attractor \cite{alberti2021small, faranda2013extreme}; or (abrupt) shifting between different basins of attraction \cite{hristopulos2021open, forgoston2018primer}, underlying major changes in the dynamics of the system. The latter shifts are often termed critical phenomena~\cite{hohenberg1977theory} or tipping points~\cite{Ditlevsen_2017} in the climate and physical sciences, while they are referred to as bifurcations leading to shifts between metastable basins of attraction in a mathematical context. Thus, the presence of noise can change in a fundamental way the physics of a complex system~\cite{hristopulos2021open}. Noise-induced phenomena also exist in the Earth system, and the possibility of high-impact climate tipping points has garnered increasing attention in recent years~\cite{Ditlevsen_2017, lenton2019climate, wunderling2021interacting}.\\ 
\indent Indeed, complex systems are multistable systems characterized by metastability, and noise usually manifests itself in a nonmonotonic dependence of the average lifetime of a metastable state on the noise intensity (or temperature). The interplay between the nonlinearity of complex systems and the stochastic fluctuations can lead to a wealth of dynamical behaviors, such as noise-induced phase transitions, emergence of ordered spatial structures~\cite{valenti2018,guarcello2013,guarcello2019,Valenti2012,Ushakov2011}, stochastic resonance~\cite{benzi1982stochastic, Spezia2008,Caccamo2019,Franzke2022,Alkhayuon2023}, stochastic resonant activation~\cite{Augello2009,Valenti2018-QNES,Grimaudo2022a}, noise-induced synchronization~\cite{yoshida2006noise}, and noise enhanced stability~\cite{Kwasniok2014,Valenti2018-QNES,Surazhevsky2021,Hastings2021}. It is worth highlighting noise-induced effects in climate change in the framework of the stochastic resonance phenomenon~\cite{Caccamo2019,Franzke2022,Alkhayuon2023}, the central role of stochastic parameterizations~\cite{Franzke2022,Palmer2019}, the scale dependent nature of noise fluctutations in chaotic turbulent systems~\cite{Biferale04, dubrulle2019beyond,alberti2023}, the non-monotonic dependence of the lifetime of transient dynamics on noise intensity in ecological systems~\cite{Hastings2021}, and the stochastic stabilization of metastable states in atmospheric dynamical regimes~\cite{Kwasniok2014}.\\
\indent Dynamical systems theory is often concerned with the study of systems that evolve deterministically over time, while statistical physics deals with the study of systems that evolve stochastically due to random fluctuations. The stochastic dynamical systems approach combines these two perspectives by modeling the deterministic evolution of a system, yet incorporating stochastic fluctuations due to environmental noise -- for example using stochastic differential equations. This overcomes the limitations of traditional modeling techniques, such as differential equations and numerical simulations, that may not capture the full complexity of physical systems. Furthermore, conventional deterministic climate models show systematic biases when compared to observed data~\cite{Palmer2019}. The complexity of the evolution of atmospheric carbon dioxide over time, for example, highlights the need to consider nondeterministic models, in which the key variables are inherent stochastic processes~\cite{Baum2022}. Furthermore, asymmetric stochastic resetting is an appealing tool to understand the fundamental features of natural disaster dynamics in different contexts, including ecosystems~\cite{Valenti2023}.\\
\indent Stochastic dynamics are relevant to many real-world systems across virtually all scales, from electrons or quarks to stars in a galaxy and galaxies in the Universe. The Earth System is no exception, and is characterized by different physical processes that act on multiple spatiotemporal scales. In particular, extreme events are characterized by the coexistence of intrinsic high dimensionality, complex nonlinear dynamics and stochasticity or environmental noise~\cite{Sapsis2021}. Furthermore, stochasticity manifests itself spontaneously in weather and climate extreme events. Indeed, stochastic weather and climate models arise because we are unable to resolve all processes and scales needed in predictive models~\cite{Sillmann2017,Evin2018,Villa2023}, and stochastic approaches can improve the match between observational data and model predictions~\cite{Franzke2015}. Finally, stochastic approaches could contribute to the evolution of artificial intelligence techniques, opening new perspectives to understand and improve the performance of weather and climate models~\cite{Palmer2019}.\\

\indent Stochastic dynamical systems theory has consequently emerged as a useful tool for understanding the behavior of complex systems that also exhibit chaotic dynamics. Dynamical indicators, such as local dimensions, persistence and Lyapunov exponents, can then be used to estimate the rate of divergence of nearby trajectories in stochastic dynamical systems. In this section, we will explore how the application of stochastic dynamical systems theory can serve as a connection between the approaches to extreme events described earlier in Sect. II and the modeling of chaotic and turbulent systems.

\subsection{Local dynamical indicators for rare events in turbulent flows}


Virtually all geophysical flows present an ubiquitous property: they are turbulent. One of the properties of geophysical turbulent flows is their chaotic and unpredictable nature which suggested to explore the connection between turbulence, chaos, and dynamical systems. In Sections \ref{sec:IIA}-\ref{sec:IIC} we have presented local dynamical systems indicators which are suitable for systems where the chaotic components dominate over turbulence. Here, we investigate how to treat systems where chaos and turbulence coexist and shape the system dynamics. 

We highlight that a {\it universal} underlying turbulent attractor has not been proven to exist for turbulent flows. A promising way to characterized highly turbulent flows in constrained geometries is the use of stochastic low-dimensional dynamical systems \cite{faranda2017stochastic}. This approach suggests that turbulent flows live on non-hyperbolic strange stochastic attractors \cite{flandoli1996random}. This property, i.e., non-hyperbolicity, is exactly strictly connected with non-homogeneity in terms of instantaneous properties of turbulent fluctuations, closely related with the universality of the statistics of fluctuations with respect to the mechanisms producing turbulence.

Turbulence, which is one of the unsolved problems of  physics, manifests itself via vortices (or eddies) of different sizes hierarchically organized in a self-similar way \cite{Kolmogorov41}. Turbulent flows are usually described in a statistical way derived from symmetry assumptions (homogeneity, stationarity, self-similarity) of the Navier-Stokes equations describing the dynamics of averaged quantities over spatial scales larger than the mean free path length of molecules \cite{Foias01}. We begin here by outlining the statistical theory that has been developed by \cite{Kolmogorov41}, based on the {\it universality} of the correlation function (or equivalently its Fourier transform, the energy spectrum) when normalized to the fluid viscosity $\nu$ and the mean energy dissipation $\langle \epsilon \rangle$, i.e., 
\begin{equation}
    S_2(\ell)\equiv \langle \delta u_\ell^2 \rangle  \propto \, \langle \epsilon \rangle^{2/3} \, \ell^{2/3},
    \label{eq:K41}
\end{equation}
where $\langle \cdots \rangle$ stands for ensemble or statistical average.
By assuming that vortices are hierarchically organized, i.e., they are derived from a cascade mechanism, the Kolmogorov universality concept should extend to higher moments field fluctuations. This means that if $\delta u_\ell$ is a measure of the velocity at scale $\ell$, then Eq. (\ref{eq:K41}) can be generalized as
\begin{equation}
    S_q(\ell) = \langle \delta u_\ell^q \rangle  \propto \langle\epsilon\rangle ^{q/3} \, \ell^{q/3}.
    \label{eq:Sq}
\end{equation}
Unfortunately, the global self-similarity assumption was one of the main failures of Kolmogorov theory, as shown by experimental evidence \cite{Parisi85,dubrulle2019beyond}. The breaking of this assumption resides in the {\it local} nature of the energy dissipation rate $\epsilon$, i.e., the phenomenon of intermittency arising from localized bursts of activity, both in time and in space. This was the starting point of the multifractal formalism developed by \cite{Parisi85}, based on the assumption that the energy dissipation rate is a statistical quantity, the probability density function (PDF) of which depends on the scale we are focusing on, i.e., 
\begin{equation}
    S_q(\ell) \propto \langle \epsilon_\ell ^{q/3}\rangle \, \ell^{q/3}. 
    \label{eq:Sqmulti}
\end{equation}
where $\epsilon_\ell$ is the energy dissipation inside a local ball of radius $\ell$.
Under the assumption of {\it local self-similarity}, i.e., $\langle \epsilon_\ell^{q/3}\rangle \propto \ell^{\mu(q/3)}$, Eq. (\ref{eq:Sqmulti}) can be written as
\begin{equation}
    S_q(\ell) \propto \ell^{\zeta(q)}
\end{equation}
where the scaling exponent $\zeta(q)=q/3+\mu(q/3)$ accounts for all possible (infinite) rescaling symmetries of the Navier-Stokes Equations, i.e., accounts for the existence of singularities in the energy cascade mechanism \cite{Benzi84,dubrulle2019beyond}. This refined theory has successfully described the statistical properties of the velocity field in experiments \cite{Muzy91,Benzi91,Biferale04,Boffetta08,Benzi08,Arneodo08}. However, this approach only provides time-averaged statistical information on velocity field fluctuations at different scales, lacking an instantaneous description of the dynamics, the latter being particularly helpful for exploring the local statistics of velocity field fluctuations \cite{dubrulle2019beyond,dubrulle2022many}. 

Starting from the non-universality of the statistics of small-scale fluctuations, \cite{alberti2023} proposed a time-dependent and scale-dependent framework for retrieving information on the symmetries of a turbulent steady state. The main idea is to overcome the limitations of previous measures of singularities (as the scaling exponents $\zeta(q)$ or the generalized dimensions $D_k$ \cite{caby2019generalized}), tracing time-averaged features of field fluctuations, by introducing two measures which are time-dependent. In this way, one answers Landau's objection to the Kolmogorov theory regarding the time and space fluctuations of the energy dissipation rate, providing an instantaneous view of scale-dependent fluctuations. The newly introduced measures can be derived by searching for extremes of the field fluctuations. We assume, for simplicity, a 1-D velocity field $u(t)$ (nonetheless, the method is suitable for any $n$-D field) which can be written as:
\begin{equation}
    u(t) = \langle u \rangle + \sum_\ell \delta u_\ell(t)
\end{equation}
where $\langle u \rangle$ is the mean field value (sometimes, the notation in turbulence is $U_0$) and $\delta u_\ell(t)$ is the fluctuating component of the velocity field at the scale $\ell$ \cite{Frisch95}. 

An additional feature associated with non-universality are intermittent phenomena. These are ubiquitous in turbulent flows and play an important role in dissipative energetics \cite{debue2018dissipation}. Intermittency can be described using the multifractal formalism ~\cite{biferale2004multifractal}, which among other things means  that energy dissipation or velocity fluctuations lives on a strange attractor with sudden, large-amplitude, low-frequency, localized bursts (i.e., rare events), induced by intermittency. An appropriate framework to deal with this behaviour is EVT, where state(time)-dependent metrics can be derived \cite{lucarini2016extremes}. We specifically seek to determine time-dependent EVT-based metrics for each velocity fluctuating component at each scale $\ell$. To this purpose, we use two metrics from EVT: the local dimension $d(t)$ and the extremal index $\theta(t)$, as introduced in Sections \ref{sec:IIA} , and \ref{sec:IIC}. In this way, instead of having a single pair $\left(d(t), \theta(t)\right)$ of descriptors of the turbulent flow we have a hierarchy of pairs, each associated with the dynamics of the fluctuating field at a scale $\ell$, i.e.,  $\left(d_\ell(t), \theta_\ell(t)\right)$. Thus, a time- and scale-dependent view of the system is obtained. 

A widely studied laboratory setup which displays the joint properties of chaoticity and turbulence is the von Karman swirling flow~\cite{zandbergen1987karman}. For such a flow under high Reynolds number turbulent conditions, the above indicators provide evidence of a scale-dependent underlying attractor whose geometric and topological properties depend on the large-scale forcing and also affect the distribution of the energy across the inertial range of scales~\cite{faranda2017stochastic}. This mirrors the role of singularities that break the global self-similarity and that can be considered the main reason of the failure to find a {\it universal} attractor for turbulent flows. These results also shed new light on the role of the inertial range of scales, where the turbulent cascade takes place, in distributing the energy injected at larger scales. Indeed, around the injection scales the individual scales are in quasi-equilibrium with low values of the local dimensions $d_\ell(t) \sim 2-3$. Conversely at smaller scales than the injection one the mean energy transfer is positive and the out-of-equilibrium energy cascade transfers energy to the small scales where viscous effects become relevant. Thus, while the statistical equilibrium at large scales is driven by a few degrees of freedom, generating a well defined low-dimensional attractor, the dynamics at scales smaller than the injection scale restore the symmetry broken by the cascade and generate a stochastic attractor. Beyond turbulent flows, this formalism can be applied to any time series and/or dynamical system. As an example, \cite{alberti2022novel} have shown that the formalism can be used to disentangle the role and the nature of noise in dynamical systems, allowing to clearly distinguish between a purely noise-like contribution, being characterized by an ergodic coverage of the available phase-space with dimensions fluctuating around 3 (as expected), and a more forcing-like contribution for a multiplicative noise, with dimensions larger than 3 which are distributed differently across the attractor.

\subsection{Noise-induced Phenomena in an Out-of-equilibrium Ecosystem}

\begin{figure}[t]
    \noindent\includegraphics[width=1.11\linewidth]{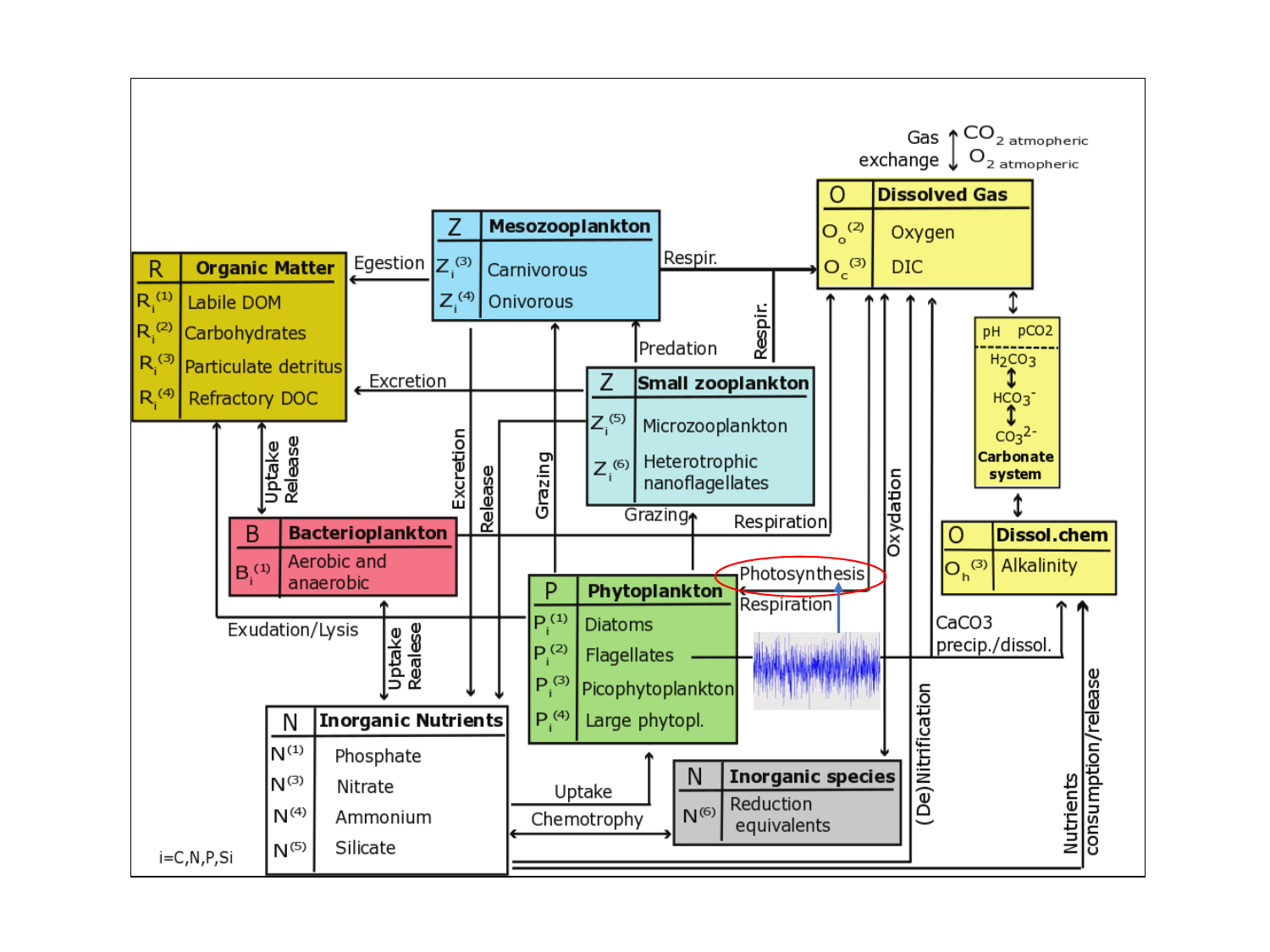}
    \caption{\textbf{Block diagram of a marine ecosystem.} The blue time series indicates the random fluctuating component of the solar irradiance, which directly affects the photosinthesis process, with effects propagating across the whole ecosystem.}
    \label{fig:Spagnolo_1}
\end{figure}

\begin{figure}
       \includegraphics[width=0.45\textwidth]{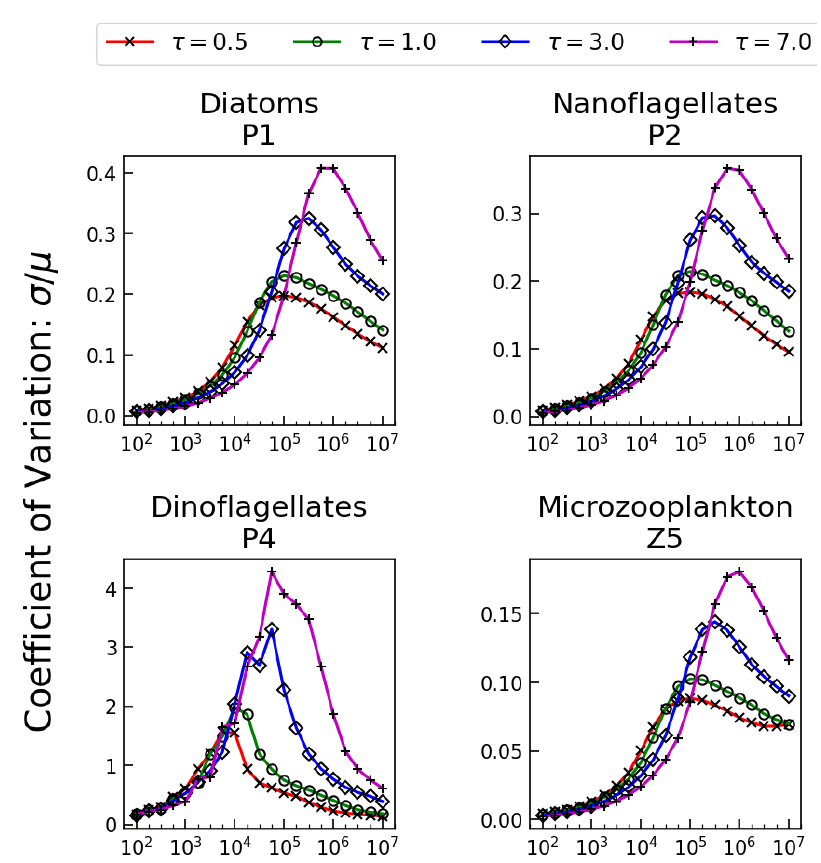}
      \caption{\textbf{Coefficient of variation versus noise intensity.} The coefficient of variation $(CV = \sigma/\mu)$ (y-axis) of four populations resulting from numerical integration of model equations plotted as a function of the noise intensity (x-axis). Different curves correspond to different values of the correlation time $\tau$.}
      \label{fig:Spagnolo_2}
  \end{figure}

Extreme geophysical events are characterized by a non-Gaussian PDF of occurrence, with rare but large fluctuations of the state variable~\cite{Sura2013}. These fluctuations, characterized by power law tails in the PDF, are peculiar of a multiplicative noise source. This means that stochastic models for geophysical extreme events with the above characteristics should be driven by multiplicative noise, with the strength of the stochastic forcing depending on the value of the state variable. For example, many relevant meteorological and climate phenomena in the atmosphere and ocean follow a non-Gaussian distribution with power-law tails and can be described by multiplicative noise forcing. This stochastic approach thus makes it possible to evaluate extreme events by understanding their underlying physics~\cite{Sura2013}.\\
Here, we present an example of how multiplicative noise plays a crucial role in modelling the dynamics of non-equilibrium ecosystems. We specifically consider the effects of randomly fluctuating solar irradiance on the population dynamics of a marine ecosystem~\cite{Grimaudo2022a}. The solar irradiance data comes from the Boussole buoy located in the Gulf of Lion, and spans 2004 to 2013~\cite{Lazzari2021b}. Fig.~\ref{fig:Spagnolo_1} shows a block diagram of the ecosystem. The blue time series indicates the random fluctuating component of the solar irradiance, which directly affects the photosinthesis process, with effects propagating across the whole ecosystem. By exploiting a $0$-dimensional stochastic biogeochemical flux model, we find a nonmonotonic behaviour of the coefficient of variation $CV$ (the ratio between the standard deviation $\sigma$ and mean $\mu$) of the marine populations’ biomass with respect to noise intensity of the solar irradiance, for different values of the noise autocorrelation time $\tau$ (see Fig.~\ref{fig:Spagnolo_2}). This indicates a noise-induced transition of the ecosystem towards an out-of-equilibrium steady state. Moreover, we see evidence of noise-induced effects on the organic carbon cycling processes underlying the food web dynamics. Different curves correspond to different values of the correlation time $\tau$. These plots show that, for a fixed value of $\tau$, there exists a value of the noise intensity for which the planktonic concentrations are maximally spread around their mean values, corresponding to the maximum value of $CV$. Furthemore, such a nonmonotonic behaviour suggests the presence of a resonance, which can be interpreted as the effect of the interplay between the nonlinearity of the system and the environmental random fluctuations. These results clearly show the profound impact that stochastic environmental variables can have on both the populations and the biogeochemistry at the basis of a marine trophic network. Non-trivial, noise-induced dynamics can push the ecosystem away from the deterministic attractor and drive it towards a new nonequilibrium steady state. This highlights the interdisciplinary link between stochastic dynamical systems theory and geophysical complex systems.

\subsection{Stochastic approaches to simulate ensembles of unprecedented events}
\label{subsect:SWG}


Data limitations (in observations and even numerical model simulations) are a major challenging in modelling geophysical extremes. This makes direct application of many mathematical analysis tools sometimes difficult to implement. Indeed, many methods rely on the assumption of an unlimited amount of data on the system (cf. Sect. \ref{sect:Define_track}). Observations of geophysical systems instead provide one finite trajectory. Even numerical simulations, such as climate model simulations, provide at most hundreds of $O(10^2)$ year-long trajectories \cite{eyring2016overview}, which limits the extremeness of the events that can be studied. The problem is made more acute by the presence of low frequency trends (including, but not limited to climate change) leading to time-scale interactions that may affect extreme events \cite{seneviratne_weather_2021-1}.

\begin{figure}[ht!]
  \noindent\includegraphics[width=0.5\textwidth]{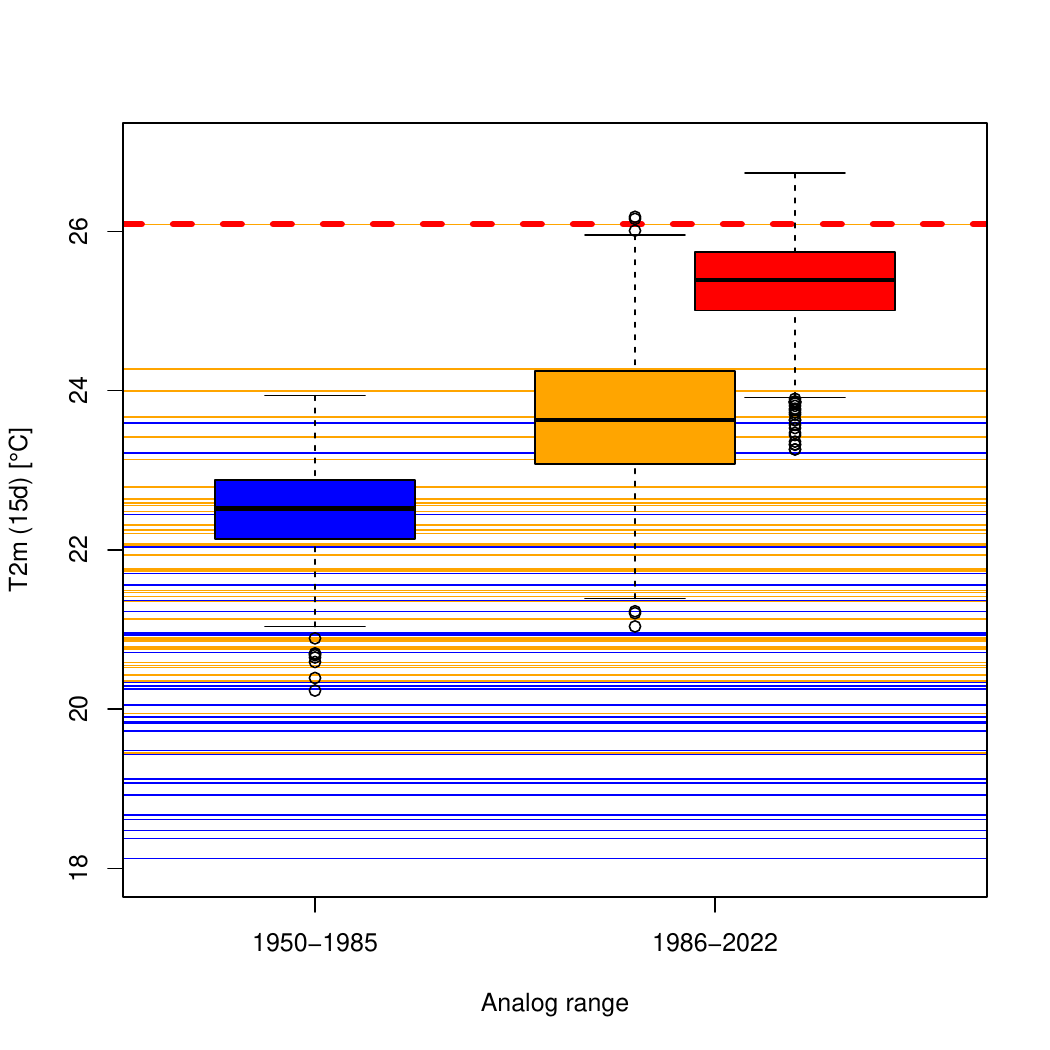}\\
  \caption{\textbf{Empirical probability distribution of extreme 15 day heatwaves in France.}  The horizontal lines represent the maximal 15 day averages of 2m temperature (in $^\circ$C)  in the summer, in France, from the ERA5 reanalysis. The blue lines are between 1950 and 1986. The orange lines are between 1986 and 2002. The red dashed line is the record value obtained in 2003.
  The boxplots represent the empirical probability distributions of 1000 simulations of the SWG with importance sampling, with an initial condition in 2003. The blue boxplot is for analogs in 1950--1985. The orange boxplot is for analogs in 1986--2022, excluding information from 2003. The red boxplot is for analogs in 1986--2022, including information from 2003.}
    \label{fig:yiou1}
\end{figure}

Record breaking events, like the European heatwave of 2003~\cite{garcia2010review} or the Western North American heatwave in 2021 \cite{lucarini2023typicality,white2023unprecedented}, have opened questions specifically focussed on the non-stationarity of the underlying system. Indeed, one can wonder how likely such extreme events are in the present climate, and how this may change in the future. We consider here two types of extreme events, where time $t$ corresponds to the publication of this study:

\begin{itemize}
\item Type 1: Events that occur at time $t$, but never occurred before $t$ (or during a reference period of observations). Hence there are documented examples and data for type 1 events, albeit small samples. 

\item Type 2: Events that have \emph{never} occurred before time $t$. By definition, no observation exists for type 2 events at $t$. Note that the 2003 European and 2021 Western North American summer heatwaves were type 2 events before 2003 and 2021, respectively. Therefore, type 2 events can become type 1 with time, as observational databases grow. 
\end{itemize}

From the  mathematical point of view, there are no fundamental differences between the two types of events, when infinite trajectories are available. The difference stems from the limited amount of “real world” observations or numerical data. We then ask the following question: can type 2 events be deduced from (known) type 1 events, and if so how?

A heuristic and pragmatic approach for simulation of type 2 events is to identify a set of initial conditions leading to intense type 1 events, and run ensembles of a climate model with perturbations around those initial conditions \cite{gessner_very_2021}. This so-called "ensemble boosting" approach can provide samples of type 2 events based on existing type 1 events. In this vein, \cite{ragone2018computation,ragone2021rare} pioneered the application of large deviation theory to the simulation of extreme European heatwaves in climate models, using importance sampling algorithms. In importance sampling, climate model trajectories are cloned or killed based on statistical rules aiming at maximising a specific observable, such as temperature. Always using large deviation theory, \cite{galfi_large_2019, lucarini2023typicality, galfi2022persistent} showed that the statistics of "long lasting" type 2 events can be inferred from type 1 events by leveraging the notion of \emph{typicality} of events for a given level of "extremeness".

As climate models yield their own biases and often underestimate climate variability and extremes, observation-based approaches are also needed to solve the type 2 anticipation challenge, given data on type 1 events. For example, \cite{yiou_anawege:_2014, yiou2020simulation} developed an observation-based stochastic model called a Stochastic Weather Generator (SWG). This SWG is a Markov chain (e.g. temperature) with hidden states (which can be the large-scale atmospheric circulation) whose transition probabilities drive the system. By modifying transition probabilities, the SWG can emulate through observations the application of rare event algorithms in a climate model \cite{yiou2020simulation}. Indeed, the SWG simulates a variable $Y(t)$ from sampling analogs of the hidden states $Z(t)$. The observable to be optimized is $f(Y)$.

Let us assume that the most extreme event (for the variable $Y$) in the "real world" occurs at time $t_x$. Then the challenge identified above can be solved by simulating the observable $f(Y)$ with analogs of the hidden variable $Z$ that include the knowledge of $t_x$ (\emph{cum data}) or \emph{not} (\emph{sine data}): can the record value of $f(Y(t_x))$ be reached or exceeded when we simulate $f(Y)$ with analogs of $Z$ that never consider information at $t_x$?

We illustrate this statistical approach on record 15 day heatwaves in France~\cite{yiou2023ensembles}. We consider the daily mean temperature averaged over France as $Y$, taken from the ERA5 reanalysis \cite{hersbach2020era5} between 1950 and 2022. We determine the warmest 15 day spells (horizontal lines in Fig. \ref{fig:yiou1}). Therefore, the observable $f$ is a 15 day average of $Y$. The record value is obtained in 2003. This value exceeds the preceding record by almost 2$^\circ$C. The hidden variable $Z$ is geopotential height at 500 hPa (Z500) over the North Atlantic region. For each day in 1950--2022, we compute the 20 best analogs of Z500 in 1950--1985 and 1986--2022. We then simulate 1000 trajectories of the SWG with importance sampling, starting in August 2003 (beginning of the hottest 15 day heatwave), using analogs in each period in turn. For the 1986--2022 period, we can exclude analogs in 2003 or include them, which allows determining the weight of this record event in the probability distribution of the simulated extremes, and evaluate to what extent the 2003 event could be anticipated without information on 2003. The result is reported in Figure \ref{fig:yiou1}.

We find that if an event is initiated in August 2003, but with meteorological conditions of 1950--1985, the most severe heatwaves cannot reach the value of 2003 (although the SWG is close to other recent heatwaves, blue boxplot). If we consider both the meteorological conditions of 1986--2022 and information from 2003, then the 2003 record can easily be exceeded (red boxplot). If information on 2003 (apart from the initial condition) are removed in the 1986--2022 SWG simulations, the 2003 value can be reached, although with low probability (orange boxplot). This means that the record-shattering event of 2003 could be anticipated. As a caveat, we note that in this example  we used information from other extreme heatwaves, three of which occurred after 2003.

\section{Applications of dynamical indicators to geophysical data}

In this section, we provide examples of a number of applications of the analysis frameworks described in Sect. \ref{sect:Define_track} to geophysical rare events. We specifically consider meridional energy transport in the atmosphere, extremes associated with persistent large-scale atmospheric circulation states typically termed "weather regimes", and spatially compouding weather extremes, namely geographically remote weather extremes that co-occur and are physically connected.

\subsection{Statistical properties, temporal and spatial scales of meridional energy transport extremes}

\begin{figure*}[ht!]
  \noindent\includegraphics[width=\textwidth]{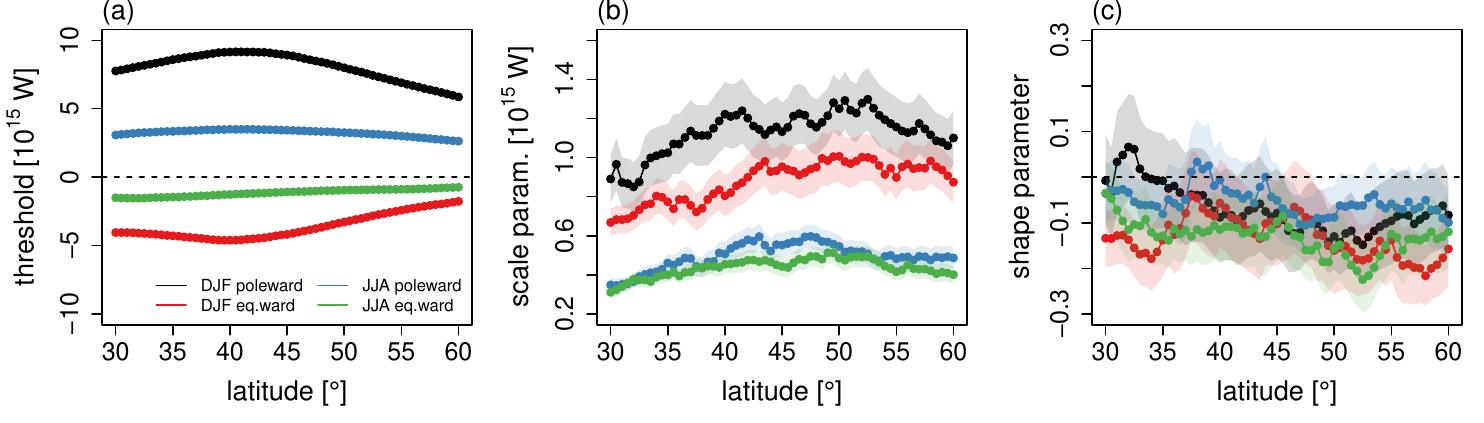}\\
  \caption{\textbf{GPD parameters for meridional atmospheric energy transport.} Meridional profiles of the GPD (a) threshold, (b) scale parameter and (c) shape parameter in case of DJF poleward (black), DJF equatorward (red), JJA poleward (blue) and JJA equatorward (green) meridional energy transport extremes. The shading represents 95\% maximum likelihood confidence intervals of the respective parameters.}
    \label{fig:lembo_1}
\end{figure*}

\begin{figure*}[ht!]
  \noindent\includegraphics[width=\textwidth]{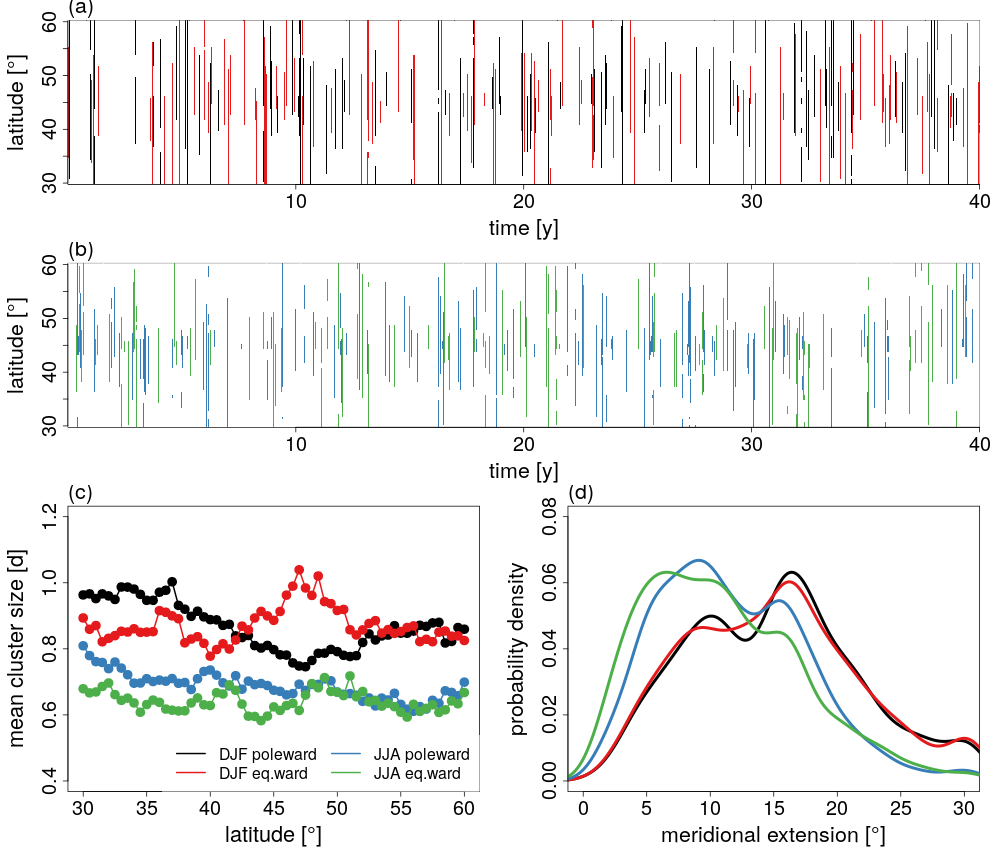}\\
  \caption{\textbf{Hovmoller plots and scale of meridional atmospheric energy transport extremes.} Hovmoller plots indicating the occurrence of: (a) DJF poleward (black) and equatorward (red) and (b) JJA poleward (blue) and equatorward (green) meridional energy transport extremes. (c) Mean cluster size and (d) probability density of the meridional extension of meridional energy transport extremes.}
\label{fig:lembo_2}
\end{figure*}

The atmosphere transports large amounts of energy meridionally, from the tropical regions to the polar regions. These meridional energy transports are intrinsically intermittent in time and space \cite{swanson1997lower,messori2013,messori2014some, messori2015,messori2017spatial,lembo2019spectral,rydsaa2021changes,lembo2022}.
We hereby review an EVT-based methodology to detect extreme events in meridional energy transports in the midlatitudes, as first presented in \cite{lembo2022}.

Meridional energy transports are obtained at every spatial point and pressure level as the scalar product of meridional velocity $v$ and total energy $E$. The total energy is given by
\begin{equation}
    E = H+ \frac{1}{2}\mathbf{v}^2 = c_p T + L_v q + gz_g
    \label{toten}
\end{equation}
where $\bf{v}$ is the horizontal velocity vector, $T$ the tropospheric air temperature, $q$ the specific humidity and $z_g$ the geopotential height, while constants $c_p$, $L_v$ and $g$ are specific heat at constant pressure, latent heat of vaporization and gravitational acceleration, respectively. The zonal and vertical mean of this quantity is obtained by spatial integration: 
\begin{equation}
    \oint{\int_{p_s}^0 vE\frac{dp}{g}dx}
    \label{spatint}
\end{equation}
where $p_s$ is surface pressure. Meridional energy transports are positive when poleward-directed, negative otherwise. We consider ERA5 Reanalysis data over 1979--2014 \cite{hersbach2020era5}, in the Northern Hemisphere. We detrend our time series and remove the seasonal and daily cycles for every latitude, and analyse boreal winter (DJF) and summer (JJA) separately.

In order to account for clusters in time that would possibly bias the estimates of the GPD parameters, we apply a declustering technique based on the above-mentioned extremal index (Eq. \ref{eq:EI}) \cite{ferro2003,galfi2017}. We estimate the extremal index based on the intervals method \cite{ferro2003} and use the obtained information to decluster the time series, thus ensuring that the selected extreme values are independent as required by EVT.

In the following, we apply EVT to both tails of the meridional energy transport distribution, namely poleward and equatorward extremes. Following \cite{lembo2022}, these are defined as events beyond the 90 \% and 10 \% percentiles of the meridional energy transport distribution. Based on the peak-over-threshold GPD scale parameters (Eq.~\ref{eq:GPD0}, ~\ref{eq:GPD}), we notice that the magnitude and variability of extremes is generally larger in winter than in summer, and it is more intense for poleward extremes than for equatorward extremes (Fig.~\ref{fig:lembo_1}a, b). The largest magnitude and variability occur for DJF poleward extremes. The threshold values for the left tails at each latitude are below zero; these are thus equatorward energy transport extremes and not weak poleward extremes. The shape parameters are predominantly negative, suggesting that meridional energy transport extremes have an upper bound (Fig.~\ref{fig:lembo_1}c). Only in the case of poleward extremes, do we observe shape parameter values close to or above zero for the latitudes south of  37 $^\circ$N in DJF, while in JJA the largest shape parameters occur between latitudes 37 $^\circ$N and 47 $^\circ$N. This suggests that, in these regions and seasons, there is a chance for “surprises”, i.e. exceptionally large poleward energy transport extremes. The shape parameter values, however, have a substantial uncertainty. Thus, we cannot exclude that shape parameters related to other types of extremes or in other regions reach values close to or above zero as well, nor that the near-zero or positive values we highlight above are indeed such.

The above-discussed GPD parameters do not provide information on the temporal or spatial scales of the energy transport extremes. To illustrate these scales, we mark the occurrence of extremes in the DJF (JJA) energy transport time series. For illustration purposes, we identify the timesteps of the extreme events using latitude 45 $^\circ$N, and then verify whether the transport attains extreme values also at other latitudes (Fig.~\ref{fig:lembo_2}a, b). We summarise the information regarding the duration and meridional extension of the extremes in Fig.~\ref{fig:lembo_2}c, d, respectively. The duration is expressed in terms of the mean cluster size, which is computed as the inverse of the extremal index. The meridional extension is given in degrees latitude, and is defined as the number of consecutive latitudes affected by an extreme event at a given time step. In case of multiple extremes occurring at a given time, marked by vertical lines with interruptions, we compute the size of the largest one. 

DJF extremes often last close to one day, and their most common meridional extension is around 17$^\circ$. Their distribution is skewed to the right, and we notice that the probability of large events extending over the full latitude band is substantially higher than the probability of small events of a few degrees latitude. JJA extremes are slightly shorter-lived -- with mean duration shorter than 1 day but longer than 0.5 days -- and smaller in meridional extent. Their distribution is skewed to the left, and the most common event has a meridional extent of approximately 9$^\circ$ in case of poleward extremes and is slightly smaller in case of equatorward extremes. The probability of events extending over the full latitudinal band is much smaller than in winter. However, the distribution has a long right tail, thus the probability of large events is not negligible. The temporal and spatial scales of meridional energy transport extremes discussed here underline their intermittent nature, and hint at their large-scale coherence and their signature in extratropical atmospheric dynamics.\\

\subsection{Identifying rare features in the large-scale atmospheric circulation associated with extreme weather events}
\label{sect:rare_gradient}

\begin{figure*}[ht!]
  \noindent\includegraphics[width=1.0\textwidth]{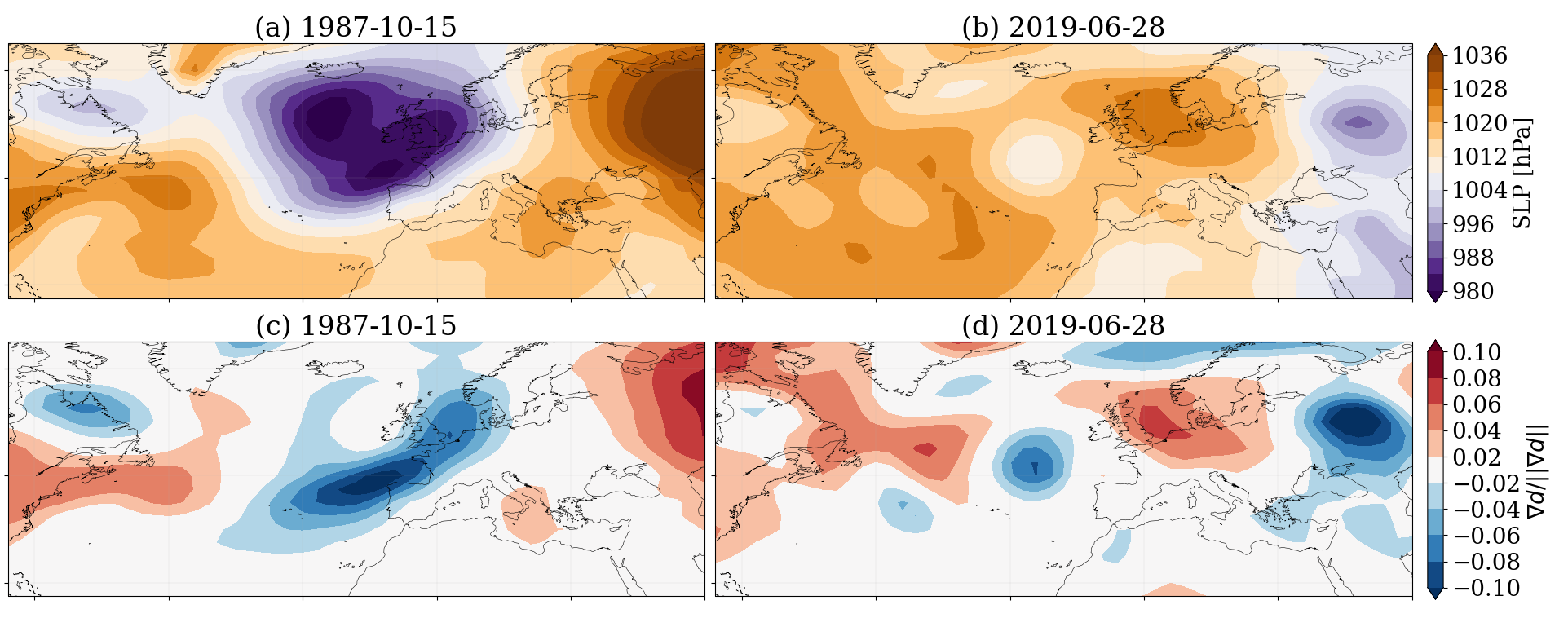}\\
  \caption{\textbf{Sea level pressure and gradient of local dimension for two extreme case studies: the 1987 Great Storm and the June 2019 French Heatwave.} SLP (hPa) for (a) the 1987 Great Storm and (b) the June 2019 French Heatwave. (c, d) Corresponding direction of the normalized gradient of local dimension on the dates indicated in the panel titles. Scalars at each grid point are components of the full vector. The figure uses NCEP/NCAR reanalysis data.} 
    \label{fig:NCEP_gradient}
\end{figure*}

We consider here the application of the estimator of the local dimensions and their gradients, discussed in Sect. \ref{sec:IIC}, to sea-level pressure (SLP) data over the Euro-Atlantic region. We specifically consider the years 1950--2020 in the NCEP/NCAR reanalysis data \cite{kalnay1996ncep} and take two extreme events as examples: the 1987 Great Storm \cite{burt1988great} and the June 2019 French heatwave \cite{mitchell2019day}. The former was a severe extratropical cyclone that affected primarily France and the United Kingdom in October 1987. The latter was a heatwave which set record high temperatures in multiple European countries, including France. The aim is to obtain information on the dynamical characteristics of the large-scale atmospheric circulation features associated with these extremes, as diagnosed through SLP. Each point $z$ in the dataset we use is defined by a daily SLP map over the chosen geographical domain. The domain for the two extremes being analysed and their SLP map is shown in Fig.~\ref{fig:NCEP_gradient}a,b.

We then compute the normalised gradient in phase space of the local dimension. At a point $z$, this is given by: $\frac{\nabla_{z} d_u(z)}{||\nabla_{z} d_u(z)||}$. This is a vector of the same dimension as $z$, and can therefore be represented on a map. The two gradient vectors for the 1987 Great Storm and the June 2019 French heatwave are shown in Fig.~\ref{fig:NCEP_gradient}c, d. Scalars at each grid point are components of the full vector. Since $\frac{\nabla_{z} d_u(z)}{||\nabla_{z} d_u(z)||}$ has an origin, which is $z$ itself, care should be taken in the interpretation of the directions in the phase space indicated by this gradient. Indeed, they must be read with respect to the point where the gradient is computed. Negative (blue) gradients mean that if one wants to increase $d$, then one should decrease the SLP values at that grid point. Positive (red) gradients means that one should increase SLP to increase $d$. As a consequence, the geographical regions where the absolute values of the normalized gradient are the largest, correspond to directions in the phase space where $d$ would change the most for a given change in SLP. In a finite data set, the geographical regions with the strongest gradients correspond to phase-space directions in which $z$ has the least number of analogues (or analogues which are further away from $z$ for a given distance metric, if a fixed number of analogues is identified). The geographical regions with very high gradients hence correspond to uncommon features of an SLP map.

Fig.~\ref{fig:NCEP_gradient}c shows that the key structure in the SLP field of the 1987 Great Storm that makes the atmospheric situation uncommon is not the low pressure situated just south of Iceland, but rather the storm itself (which also corresponds to a low pressure) situated south-east of Brittany. Similarly for the June 2019 French heatwave, Fig.~\ref{fig:NCEP_gradient}d shows that the atmospheric feature which determines the rarity of the situation is the low pressure situated above Russia. The low pressure to the east of Ireland and the high pressure in Northern Europe are also uncommon atmospheric features.

For each point in the phase space of some geophysical system, the gradient of the local dimension thus indicates structures which are rare or common with respect to the sampling done by the data set. We foresee that investigating the link between this gradient and the dynamics of the system may be a fruitful future line of research.

\subsection{From transition probabilities to the predictability of midlatitude extreme weather}


Atmospheric motions are chaotic, which implies an unavoidable loss of predictability with increasing lead time. However, in chaotic systems it is possible to identify recurrent patterns. These patterns have often been conceptualized as so-called weather regimes, namely clusters of similar atmospheric states \cite{michelangeli1995weather}. Weather regimes are then recurrent, quasi-stationary, and persistent large-scale atmospheric circulation states \cite{hannachi2017low} usually defined as typical clusters of atmospheric flows that are observed in an specific geographical region. Weather regimes appear as sticky regions of the phase-space where the trajectories slow down, possible due to the vicinity of stationary or quasi-stationary solutions (e.g. \cite{michelangeli1995weather,hochman2021atlantic}).

\begin{figure*}[ht!]
\noindent\includegraphics[width=1.0\textwidth]{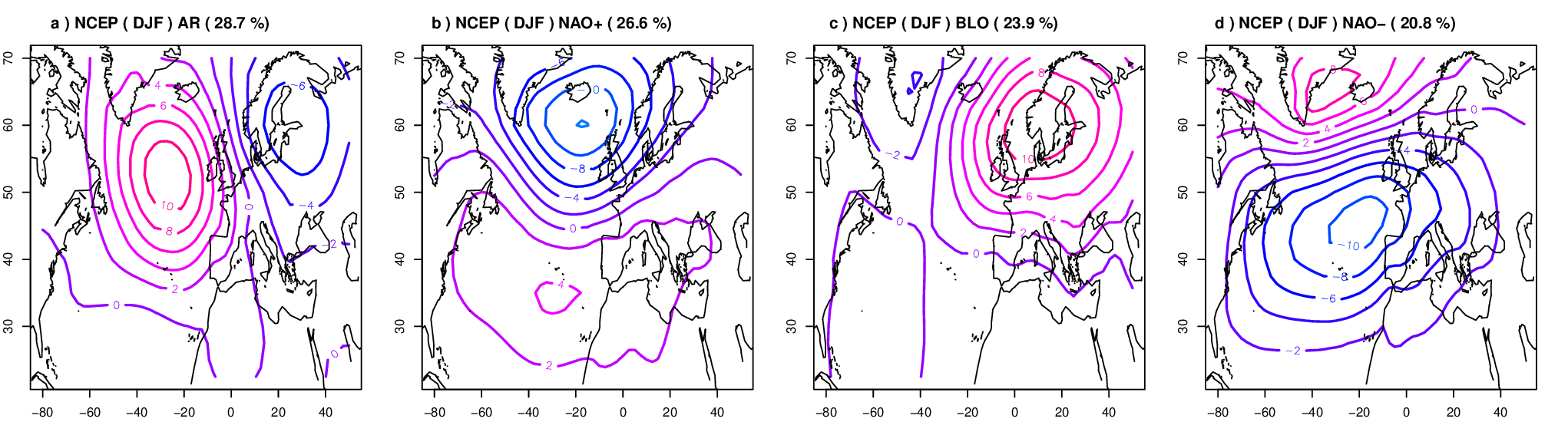}
\caption{\textbf{Euro-Atlantic Weather regimes}. Weather regimes in the Euro-Atlantic sector [80ºW-50Eº, 20º-70ºN] computed using sea level Pressure from NCEP/NCAR reanalysis \cite{kalnay1996ncep} over 1948--2021, following the methodology of \cite{yiou2004extreme}.}\label{Alvarez1}
\end{figure*}

Weather regimes have been defined over a number of different geographical regions, but their use has been most widespread in the Euro-Atlantic sector. There, the optimal number of weather regimes is typically taken to be four~\cite{vautard1990multiple}, although fewer \cite{dorrington2022quantifying} or more \cite{grams2017balancing} have been proposed. Figure \ref{Alvarez1} shows the four canonical regimes using sea-level pressure (SLP) over the region [80ºW-50Eº, 20º-70ºN]. These are, ordered by decreasing frequency of occurrence: 

(a) the Atlantic Ridge (AR), with a high pressure anomaly over the center of the North Atlantic;
(b) the positive phase of North-Atlantic Oscillation (NAO+) with a dipole of anomalously low pressure in the northern North Atlantic and anomalously high pressure to the south;
(c) Scandinavian Blocking (BLO), with an anomalous high pressure center over Scandinavia;
(d) the negative phase of North-Atlantic Oscillation (NAO-), showing a dipole which is roughly inverse to that of the NAO+.

Weather regimes can establish conditions favorable for large-scale extremes (\cite{yiou2004extreme, lavaysse2018use,beerli2019stratospheric,pasquier2019modulation,ruggieri2020north,hauser2023linkage,alvarez2018atmospheric}), and are used in the context of medium-range and sub-seasonal to seasonal weather forecasts and climate projections (\cite{cortesi2019characterization, cortesi2021yearly, merryfield2020current}). This motivates an interest in studying the large-scale atmospheric evolution during extreme weather events from a weather regimes perspective.

\begin{figure}[ht!]
 \centerline{\includegraphics[width=23pc]{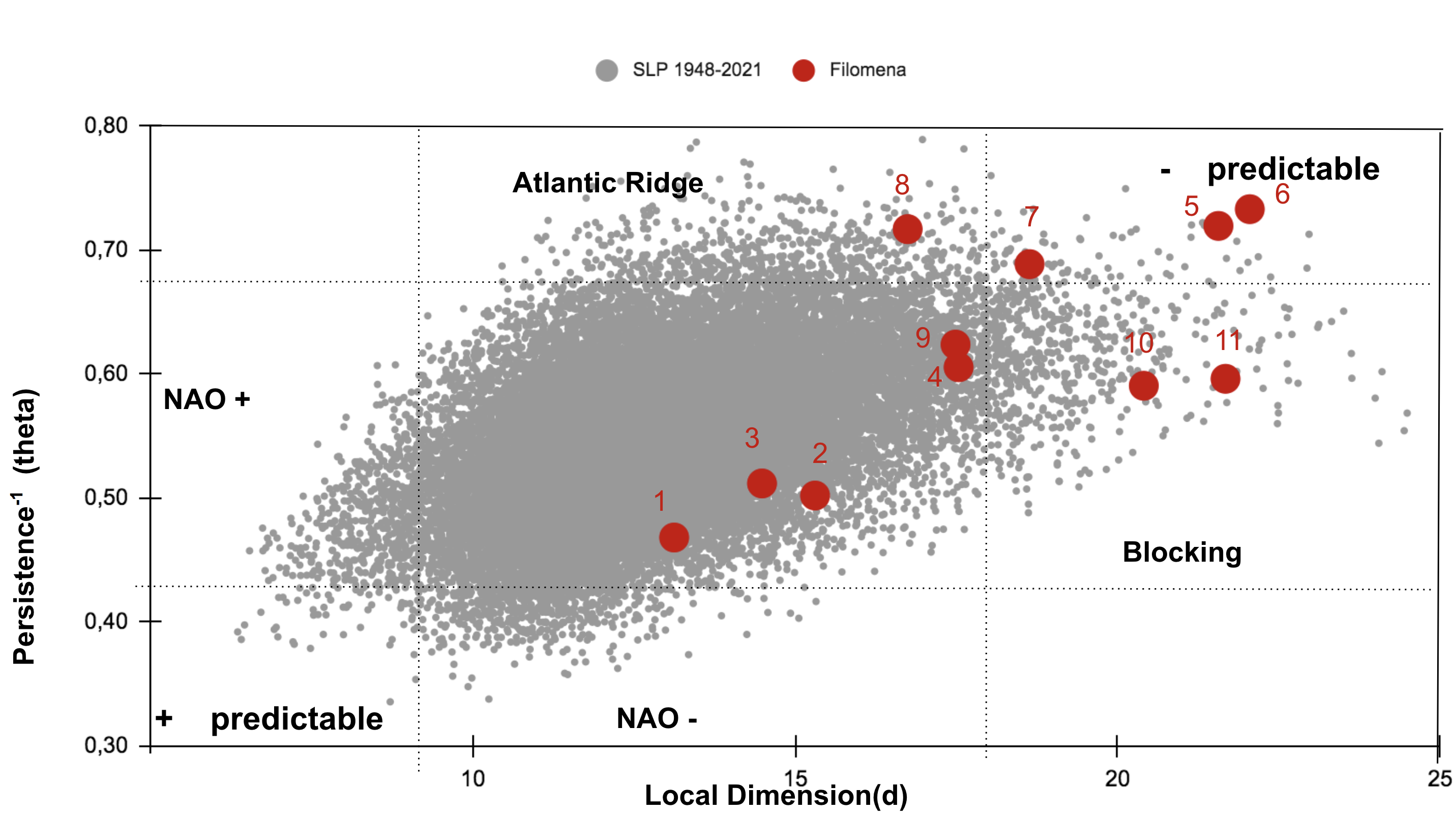}}
 \caption{\textbf{A dynamical systems and weather regime analysis of storm Filomena}. The grey points in the scatterplot show daily $d$ and $\theta$ values computed using SLP from the NCEP/NCAR reanalysis over the period 1948--2021, in the region [80ºW-50Eº, 20º-70ºN]. Red dots show  the values during storm Filomena, numbered consecutively starting from the $5^{th}$ of January 2021. The weather regime labels in the different quadrants of the plot are based on \cite{faranda2017dynamical}. }\label{Alvarez3}
\end{figure}

The weather regimes and their transitions have also been connected to the dynamics of the underlying atmospheric system. \cite{faranda2017dynamical}, showed a statistical link between anomalies in $d$ and $\theta$ computed on the SLP in the Euro-Atlantic sector and the four canonical weather regimes (see labels in Fig. \ref{Alvarez3} here). Later, \cite{hochman2021atlantic, platzer2023dynamical,lee2024dynamical} found that $d$ and $\theta$ reflect the life cycles of both Euro-Atlantic and North American regimes. Specifically, $d$ and $\theta$ decrease when a regime is established, and increase during transitions between different regimes, albeit with some regime dependence.

The two dynamical systems metrics have also been connected to the intrinsic predictability of the atmosphere. The argument is that low-dimension, high persistence atmospheric patterns should afford a higher predictability than high-dimension, low persistence cases \cite{faranda2017dynamical, messori2017dynamical, scher2018predicting, hochman2021new, hochman2022dynamics}, although this may not always directly map to the performance of numerical weather forecasts \cite{scher2018predicting}.

We illustrate here how combining an analysis based on $d$ and $\theta$ as described in Sect. \ref{sec:IIC} with one of regime transitions can shed light on weather extremes. As example, we pick storm Filomena (Fig. \ref{Alvarez3}): a storm which affected the Iberian Peninsula in January 2021, bringing heavy snowfall to large regions of Spain \cite{tapiador2021satellite,zschenderlein2022unusually}.

The first days of storm Filomena corresponded to an NAO- type regime regime. This transitioned very rapidly to a rare atmospheric configuration, which was very unusual in terms of $d$ and $\theta$ values and which did not correspond closely to any of the four canonical weather regimes. The very low persistence (high $\theta$) and high $d$ on these days suggests that the atmosphere was in a low predictability situation. It was during these days that intense snowfall occurred in Spain. Next, the atmosphere transitioned towards AR and from AR to BLO.  

The example of Filomena shows that rare and low-predictability weather regimes as diagnosed by $d$ and $\theta$ can be associated with extreme events. This was also noted by \cite{hochman2019new} when investigating snowfall events in the middle-east. This type of analysis in $d$--$\theta$ space can be applied to a range of different extreme event categories, as a complement to a categorical study of weather regimes and their transitions. Indeed, the case of Filomena shows that the circulation associated with extremes does not always fit neatly in one of the canonical regimes. The proposed analysis also informs on the temporal evolution of the large-scale atmospheric predictability during the extreme event from a dynamical systems perspective. 

\subsection{Co-recurrence statistics for spatially compounding weather extremes}\label{sect:Compound}


Understanding the drivers and dynamics of weather extremes is facilitated by a multivariate analysis framework. Indeed, different weather extremes are often associated to common physical drivers (for example a cyclone causing both strong winds and heavy rain), and even geographically remote extremes can be physically connected \cite[e.g.][]{messori2016cold} -- so-called spatially compounding extremes. Even extreme events that relate to a single impact variable benefit from a multivariate analysis. They can be associated with multiple drivers (e.g. a heatwave may be driven by both the atmospheric circulation and soil moisture characteristics \cite{fischer2007soil}), or require investigating the impact variable and a possible driving variable jointly.

Here, we consider both a monovariate extreme and the case of spatially compounding extremes. Such extremes can pose a greater threat to human societies than single, isolated extremes, as their effects may exacerbate each other and lead to correlated losses \cite[e.g.][]{mills_insurance_2005, kornhuber2020amplified}. We specifically illustrate the application of multivariate dynamical systems indicators to first study drivers of cold spells in North America and next their co-occurrence with cold spells in North America and wet or windy extremes in Europe -- sets of extreme weather events which have been highlighted in the literature as being spatially compounding \cite{messori2016cold, leeding2023pan,leeding2023modulation,messori2023systematic,riboldi2023multiple}.

\begin{figure}[ht]
\noindent\includegraphics[width=0.5\textwidth]{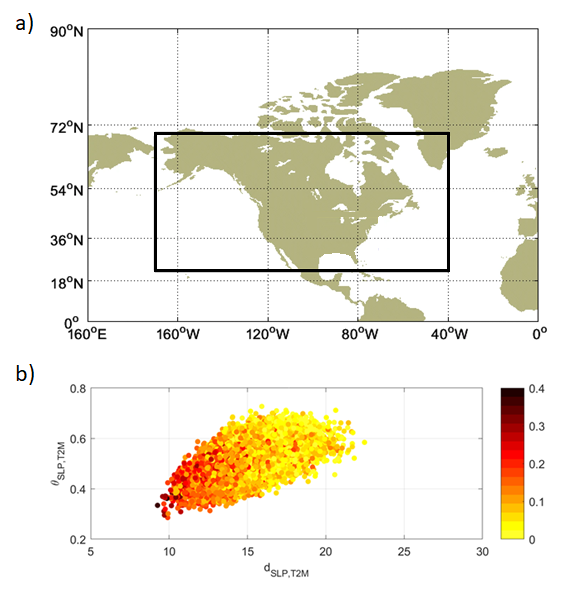}
 \caption{\textbf{Geographical domain and scatter plot for the dynamical systems indicators computed on SLP and t2m.} a) The black rectangle highlights the domain over which the dynamical systems metrics are computed. b) $d$--$\theta$ scatter plot coloured with the values of the co-recurrence ratio $\alpha$. The figure is adapted from \cite{faranda2020diagnosing} and uses NCEP/NCAR reanalysis data \cite{kalnay1996ncep} over 1948--2018.}\label{N_Am_scatter}
\end{figure}

We implement here the previously discussed estimation of $d$, $\theta$ and $\alpha$ in a bivariate context (Sect. \ref{sec:IIC}).
We illustrate in Fig. \ref{N_Am_scatter} these indicators computed for SLP and 2-meter temperature (t2m) over North America. The atmospheric configurations with a high co-recurrence ratio $\alpha$ between SLP and t2m tend to be those with relatively low $d$ and $\theta$. When a given pair of SLP and t2m patterns recurs, it thus tends to coincide with low-dimensional and persistent configurations. 

\begin{figure}[ht!]
 \noindent\includegraphics[width=0.3\textwidth]{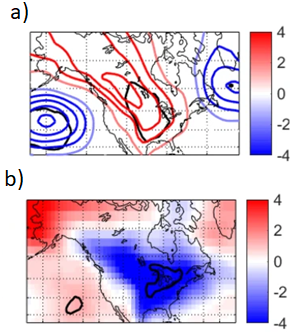}
 \caption{\textbf{Composite SLP and t2m anomalies for winter days with a high co-recurrence ratio.} Composite SLP anomalies in hPa (a) and t2m anomalies in K (b) corresponding to the 10\% most positive anomalies of $\alpha$ relative to its seasonal cycle, during DJF. The black lines bound regions where at least 2/3 of the composited anomalies have the same sign. The figure is adapted from \cite{faranda2020diagnosing} and uses NCEP/NCAR reanalysis data \cite{kalnay1996ncep} over 1948-–2018.}\label{N_Am_cold}
\end{figure}

We next focus specifically on days displaying anomalously high $\alpha$ in winter, and look at the corresponding SLP and t2m anomalies (Fig. \ref{N_Am_cold}) -- the focus of our analysis here. High co-recurrence ratio days display a tripolar SLP anomaly pattern, favouring the meridional advection of cold air and below average temperatures over a large part of the North American continent. Coupled with the information provided by Fig. \ref{N_Am_scatter}b, this points to cold wintertime spells being favoured by persistent circulation patterns, and to the fact that whenever cold days occur, similar large-scale t2m and SLP patterns are found. This connects back to the concept of large-scale meteorological patterns, whereby specific regional climate extremes are associated with recurrent atmospheric configurations \cite[e.g.][]{cellitti2006extreme, grotjahn2016north} and, albeit qualitatively, to the notion of typicality (See Sect. IIIc) .

\begin{figure*}[ht!]
 \noindent\includegraphics[width=0.8\textwidth]{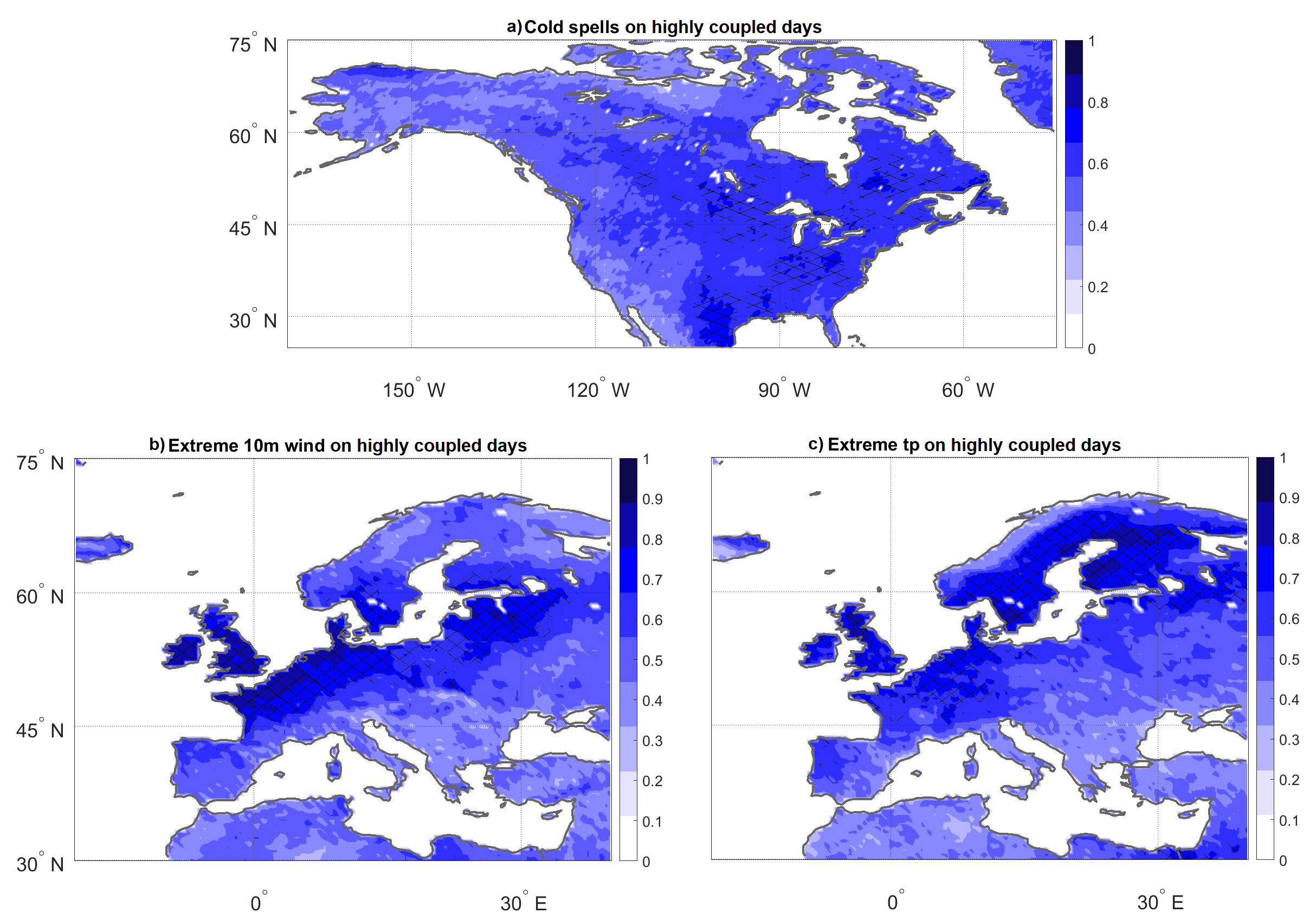}
 \caption{\textbf{Normalised occurrence of extreme climate events for winter days with a high co-recurrence ratio.} Normalised occurrence of (a) cold spells, (b) extreme 10-metre wind and (c) extreme precipitation during the 50 winter days with the highest co-recurrence ratio between SLPs over North America and Europe.
Cross-hatching marks regions where the occurrence of extremes during high co-recurrence days is significantly higher than climatology at the 5\% one-sided level, determined using 1000
random sampling iterations. The figure is taken from \cite{messori2023systematic} and uses ERA5 reanalysis data \cite{hersbach2020era5} over 1979--2020.}\label{N_Am_cold_Eur}
\end{figure*}

We next consider how the co-recurrence ratio can provide information on spatially compounding climate extremes. We again analyse wintertime cold spells over North America, but now investigate their connection to wet or windy European extremes. We specifically compute the co-recurrence ratio between SLP over North America and Europe. We then select the 50 winter days with the highest co-recurrence ratio and investigate whether they correspond to local climate extremes (defined here as t2m temperature anomalies below the local 5th percentiles and precipitation and 10-metre wind anomalies above the local 95th percentiles). There is a clear signal of heightened frequency of cold extremes over south-eastern North America and wet or windy extremes over western-continental Europe on high-coupling days (Fig. \ref{N_Am_cold_Eur}). This supports previous literature arguing for a systematic physical and statistical connection between these sets of extremes \cite{messori2016cold, de2020compound, leeding2023pan,leeding2023modulation,riboldi2023multiple}. We thus conclude that the application of the above dynamical systems metrics in a bivariate context can provide useful information on compound climate extremes. We foresee that a full multivariate implementation could be profitably applied to the study of a range of different geophysical systems.\\

\section{Opportunities and open Challenges for the study of geophysical rare events}

While several advances have been made in the understanding, modeling and simulation of geophysical extreme or rare events, several challenges still limit our capabilities of modeling small-scale or sporadic extreme events. Two examples of geophysical extreme events having these characteristics are heavy thunderstorms and tropical cyclones (TCs). Both are highly localized convective events, which are affected by physical processes on a large range of spatial scales. This makes it challenging to monitor and predict their occurrence.\\

A final challenge in studying all these geophysical extremes is the impact of climate change, which can alter the frequency, duration, and intensity of these events (see also Sect. \ref{subsect:SWG}). In this perspective, a growing attention has been devoted to so-called "attribution" studies, namely determining whether a specific extreme geophysical event may be ascribed to anthropogenically-forced climate change \cite{national2016attribution}. Predicting the impact of climate change requires the integration of non-stationarity in statistical and dynamical models, to separate the role of forced versus natural variability in the occurrence of these phenomena. This challenge is contextualised mathematically in Sect. V.D. \\

In this section we discuss in more detail the recent advances and challenges in studying the above-discussed geophysical extremes, with a focus on the opportunities issuing from dynamical systems theory and statistical physics, not least for extreme event attribution. We specifically outline lines of research that have been made possible by recent findings, but are yet to be explored in the literature.

\subsection{Physics and dynamics of convective precipitation phenomena}

Convective events are some of the most intense precipitation events on the planet, associated with rising plumes of moist air in response to instability in the atmosphere. Convection triggers updraft and downdraft and as a result is the main process of vertical exchange of water vapour, heat and chemical species between the lower and upper troposphere.\\

Convection plays an important role in both the mid-latitudes and the tropics. In the mid-latitudes it can give rise to heavy thunderstorms; in the tropics, it is a key ingredient of TCs. TCs are synoptic-scale ($O(10^3)$ km) vortices fueled by the thermal energy accumulated in the ocean. This is transferred to the atmosphere through evaporation and convective processes, and finally released through precipitation and partly converted into kinetic energy (extreme winds). Specifically, the aggregation of convective towers, or clusters of thunderstorms, can lead to the formation of a TC as the energy released from the towers fuels the development of a low-pressure center, which can then intensify into a cyclonic system.\\

The study of thunderstorms and TCs is an active area of research, with significant progress having been made in understanding the physical processes that drive these events. Examples include the role of atmospheric instability in the development of thunderstorms~\cite{radler2019frequency} and the role of large-scale atmospheric circulation patterns in the formation and intensification of thunderstorms and TCs e.g. ~\cite{tippett2012association,camargo2016tropical,faranda2022climate}. A further area of research is the role of kinetic energy, moisture and heat budgets in the development of thunderstorms and TCs. These budgets describe the exchanges of energy and moisture between the atmosphere and the surface, span a continuum of spatial scales and are important for understanding how storms and TCs develop and intensify \cite{faranda2018computation,dubrulle2022many}. We begin here by considering statistical physics approaches to the study of energy tranfers in convective systems.\\

Diagnostics of inertial energy transfers typically rely on filtering approaches that separate resolved fields from subfilter-scale fields \cite{leonard1975energy}. Filtering is broadly used in the study of turbulent geophysical flows, including both atmospheric and oceanographic flows \cite{germano_1992, meneveau_statistics_1994, tong_experimental_1999, vercauteren_subgrid-scale_2008,vannitsem2017evidence}. In all these studies, a filter is applied to the Navier-Stokes equation, and this is used to infer the energy transfers from the reference filtered lengthscale $\ell$ to larger and smaller scales arising due to non-linear interactions \cite{meneveau_statistics_1994}. Using this approach, \cite{sroka_organized_2021} diagnosed organised regions of upscale and downscale inertial energy transfers in the hurricane boundary layer, based on remotely sensed wind observations during Hurricane Rita (2005).\\

In an alternative approach based on the weak solution formalism, \cite{duchon2000inertial} showed that energy transfers in a fluid at an arbitrary scale $\ell$ satisfy a local energy balance equation:
\begin{widetext}
 \begin{equation}
\partial_t E^\ell + \partial_j \left(u_jE^\ell+\frac{1}{2}\left(u_j\hat{p}+\hat{u}_j p\right)+\frac{1}{4}\left(\widehat{u^2 u_j}-\widehat{u^2}u_j\right)-\nu\partial_j E^\ell \right)= -\nu\partial_j u_i\partial_j \hat{u}_i -\mathscr{D_\ell},
\label{DREbalance}
\end{equation}
\end{widetext}
 where $u_i$ are the components of the velocity field, $p$ the pressure, $\hat{u}$ and $\hat{p}$ their coarse-grained components at scale $\ell$, and $E^\ell=\sum_i \frac{\hat{u}_iu_i}{2}$ the  kinetic energy per unit mass at scale $\ell$ (such that $\lim_{\ell\to 0} E^\ell=u^2/2$). The term $\mathscr{D_\ell}$ is expressed in terms of velocity increments $\delta \vec u (\vec r,\vec x) \overset{def}{=} \vec u(\vec x + \vec r) - \vec u(\vec x) \equiv \delta \vec u (\vec r)$ (the dependence on $\ell$ and $\vec x$ is kept implicit) as:

\begin{equation}
\mathscr{D}_\ell (\vec u) = \frac{1}{4\ell} \int_\mathcal{V} d\vec r \ (\vec\nabla G_\ell)(\vec r) \cdot \delta\vec u(\vec r) \ |\delta\vec u (\vec r)|^2.
\label{DRfieldnotGeneral}
\end{equation}
In this definition, $G$ is a smooth filtering function, non-negative, spatially localized and such that $\int d\vec r \ G(\vec r)=1$, and $\int d\vec r \ \vert \vec r \vert ^2 G(\vec r) \approx 1$. The function $G_\ell$ is rescaled for a given volume with $\ell$ as $G_\ell (\vec r) = \ell^{-3}G(\vec r/\ell)$.\\

The 2D filter is adapted from \cite{kuzzay2017new} as a circular symmetric filtering function of the scalar increment $r$ given by 
\begin{equation} \label{eq:G}
         G(r)=
    \begin{cases}
      \frac{1}{N}\exp(-\frac{1}{1-(r/2a)^2}), & \text{if } r<2a, \\
      0, & \text{otherwise}.
    \end{cases}
\end{equation}
where $N$ is a normalization constant such that $\int d^2r G(r)=1$\\

\begin{figure*}[ht!]
\noindent\includegraphics[width=1.0\textwidth]{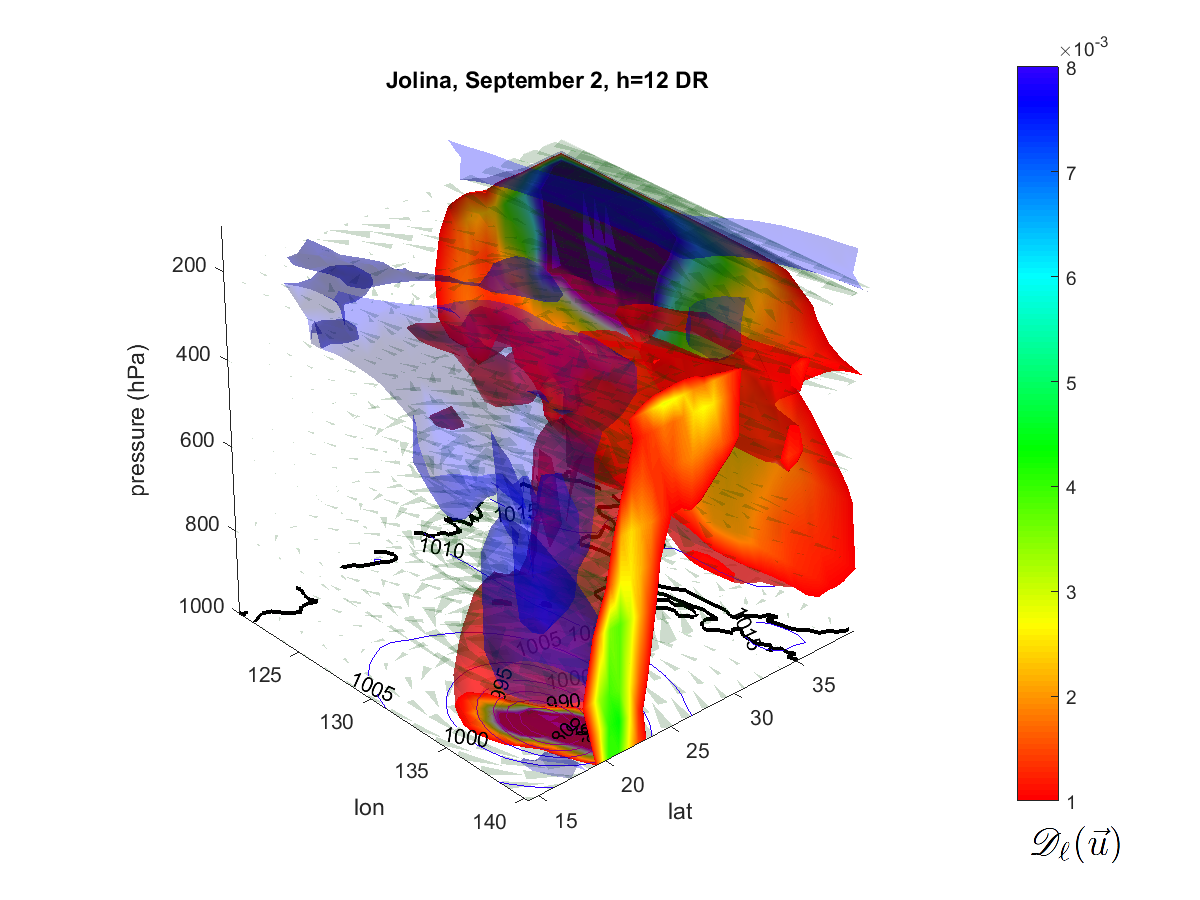}
\caption{\textbf{Mesoscale to convective scale energy transfers in Typhoon Jolina}. Energy transfer $\mathscr{D}_\ell$, with $\ell=250$ km computed using the 3D velocity fields from the NCEP/NCAR reanalysis dataset for cyclone Jolina (2nd September 2005, 12h UTC, when the cyclone reached peak intensity). The colorscale indicates energy transfers directed towards smaller (red) or larger (blue) scales than $\ell$.}
\label{fig:dujolina}
\end{figure*}

We provide here an example of applying the framework from \cite{duchon2000inertial} to the spatial distribution of instantaneous energy transfers in TCs. We specifically consider Typhoon Jolina, which hit Japan in September 2005. We use the NCEP/NCAR reanaylsis data~\cite{kalnay1996ncep} and a scale $\ell=250$ km. Positive $\mathscr{D}_\ell$ indicate transfers from the mesoscale ($\simeq$250 km) towards the smaller convective scales, while negative values correspond to tranfers towards the larger synoptic scales. Fig. \ref{fig:dujolina} suggests that energy transfers are organized around the eye of the cyclone and that intense downscale energy transfers occur in a relatively small region of the planetary boundary layer, where hazards such as heavy rainfall and wind-gusts are also concentrated. However, both the direct and inverse energy cascades coexist, and transfers towards the synoptic scales are found in cyclone's outflow in the upper troposphere. Following such energy transfers over time along the trajectory of TCs  would offer a way to study, in scale spaces, rapid intensification and rapid weakening phenomena. The above approach could also be used to investigate other convective events such as medicanes, mesocyclones, squall lines or derechoes. However, such systematic investigations are still absent from the literature.\\

\subsection{Statistics and modelling of convective precipitation phenomena}

A complementary approach to studying the energy cascades in individual TCs is to take a bulk view of cyclones as point vortices and investigate the environmental conditions that favour or suppress their growth. In fact, TCs can be viewed as a type of heat engine, where the warm ocean surface provides the fuel for the storm to extract energy and convert it into wind. Despite the advances in understanding the dynamics of individual TCs, key open questions remain on TC bulk statistics. These include:

\begin{enumerate}
    \item What controls the yearly number of observed TCs (global rate of about 70--90 events per year~\cite{sobel2021tropical})?
    \item What is the response of TC activity to variations in the mean state of climate?
\end{enumerate} 

Using a bulk perspective, \cite{emanuel1986air} developed the potential intensity theory, which predicts the maximum possible wind speed that a TC can attain based on the underlying ocean and atmospheric conditions. This theory has represented a step forward in understanding the frequency of TCs, and has led to the development of a number of semi-empirical indices know as Genesis Potential Indices~\cite{camargo2007tropical,emanuel2004tropical, tippett2011poisson, wang2020dynamic, menkes2012comparison,cavicchia2023tropical}).
However, GPIs only partly address question 1 above, since they do not take into account the small-scale dynamics leading to the “seeding”. In other words, the process or family of processes responsible for the formation of the precursor disturbances, from which self-organized convection and then the cyclones originate \cite{emanuel2022tropical}. Indeed, the actual number of TCs is only partially accounted for by the large-scale factors entering the GPI, the remaining part depending on the amount of seeds~\cite{emanuel2022tropical}. The GPI approach also struggles in addressing question 2. Indeed, when using high-resolution climate model simulations to estimate future TC activity, it is found that the trends of GPI are not consistent with those of simulated cyclones, often disagreeing even in the sign of change \cite{cavicchia2023tropical}.\\

These limitations have motivated new lines of research, such as synthetic tracking. This involves using dynamical systems and statistical methods to generate large numbers of virtual TCs, which are then used to study the bulk statistical properties of the cyclones \cite[e.g.][]{emanuel2006application,emmanuel2017using}. Specifically, a combination of deterministic and stochastic approaches are used to iteratively test "seeding" cyclone cores at different locations with different environmental conditions and then identifying how different possible tracks affect the development of the cores. Synthetic tracking has proven to be a valuable tool for predicting the paths and impacts of TCs, which can have significant implications for disaster planning and response efforts \cite[e.g.][]{hallegatte2007use,leijnse2022generating}. The very large number of tracks generated by synthetic tracking algorithms has supported a better understanding of the global rate of TC occurrence, for example by informing on the likelihood  of TC occurrences in regions that may not have experienced any in the historical period \cite{rumpf2007stochastic, bloemendaal2020generation}. Similarly, it has aided in exploring the implications of climate change for TC frequency and characteristics~\cite{walsh2016tropical,lee2020statistical}. These points are of direct relevance to the two questions listed above.\\

In addition to TCs, thunderstorms and other convective precipitation phenomena have also been also been investigated leveraging concepts from dynamical systems theory and statistical physics. An example is a framework for predicting the onset and evolution of thunderstorms using large scale environmental conditions~\cite{tippett2012association,frenkel2012using,allen2015empirical}. This approach involves using statistical mechanics to derive probabilistic models of the atmospheric processes that govern the behavior of thunderstorms, including the interaction between convection and the large-scale atmospheric circulation. Another approach that has been used to study thunderstorms is the application of Ising models~\cite{khouider2010stochastic}. Ising models are statistical physics models that are typically used to describe the behavior of interacting spins in a lattice. However, they have been adapted to describe the atmosphere as a grid of interacting variables, with each variable representing a particular atmospheric state or parameter. By analyzing the statistical properties of the variables and their interactions, researchers can gain insights into the behavior of thunderstorms, including the onset, growth, and decay of convective activity. While Ising models are a relatively new approach to studying thunderstorms, they have shown promise in capturing some of the complex dynamics of these storms and could potentially lead to more accurate predictions of their behavior in the future~\cite{hottovy2015spatiotemporal}.\\




Modelling convection presents several challenges. Perhaps the most pressing is the scale mismatch between the small scales at which convection occurs and the larger scales that can be directly resolved by numerical climate models. To address this challenge, parameterization schemes are used to represent the effects of convection on larger scales. However, these schemes are a major source of uncertainty in climate simulations. Convection also involves interactions between the atmosphere, the surface, and other physical processes such as radiation and moisture transport. Additionally, the behaviour of clouds and their evolution is affected by factors such as aerosols and atmospheric dynamics, which further complicates the modelling process. Accurately capturing all of these processes precludes a clear separation of scales as often performed in the study of stochastic dynamical systems. In fact, while ad-hoc studies employing scale decomposition approaches are possible \cite{duchon2000inertial, faranda2018computation, alberti2021small}, they cannot account for the full range of processes leading to convective precipitation.\\


A possible fashion to address this challenge, and reduce the uncertainties associated with parametrisation schemes, is to develop numerical climate models with km-scale resolution, akin to the latest generation of numerical weather prediction models. These allow to directly resolve the small scales of convection. Both single-model and multi-model ensembles of such models are important for improving robustness of results, rather than relying on individual model simulations. Single-model ensembles involve running multiple simulations with slightly different initial or boundary conditions, or with slightly different model parameter settings, while multi-model ensembles refer to  running different models in comparable set-ups. These approaches allow us to account for the uncertainty inherent in the climate system's chaotic evolution and reduce the uncertainty associated with model structure and parameters. Over the past years, several coordinated efforts have emerged to develop such ensembles (e.g. \cite{ban2014evaluation,coppola2020first,fosser2020convection,pichelli2021first}).\\

\subsection{Dynamical systems attribution of extreme weather events under climate change}

Attribution studies seek to answer the question: how has human-induced climate change modulated the probability and/or physical characteristics of a given extreme event? 
A wide range of statistical and modelling tools have been used in attribution studies, including numerous different modelling and statistical approaches \cite{national2016attribution}. Relatively recently, dynamical systems theory has also been used in attribution studies~\cite{faranda2022climate, ginesta2023methodology}.  Since the atmosphere is a complex system whose dynamics is influenced by many factors, including the ocean, land surface, and human activities, dynamical systems theory provides a framework for understanding how changes in one part of the system can affect the behavior of the entire system, including the occurrence of extreme weather events. A key concept in dynamical systems theory is the use of "recurrences" (see Sect. \ref{sect:Define_track}). This involves identifying weather patterns similar to those associated with a given event, but occurring at a different time. To study the influence of climate change on a specific event (e.g. a heatwave), one can select analogues for the circulation (e.g. sea-level pressure maps) associated to that event in past and present periods in an observational-based dataset. Differences between present and past analogues can enable to make an attribution statement. Going back to our example, if analogues of a heatwave in a past period results in significantly cooler near-surface temperatures than those associated with the analogues in the present period, this may be used to argue for a measurable impact of anthropogenic climate change on the event. 
Dynamical systems theory therefore helps to condition attribution to a specific circulation. 
A related approach can also support diagnosing the physical processes associated with the attribution. For example, if there is a persistent atmospheric circulation pattern that is associated with extreme weather events, such as a blocking high-pressure system, then researchers can use dynamical systems theory to understand how changes in the climate system can affect the persistence and intensity of the circulation pattern, and how this can modulate the associated extreme weather events.~\cite{faranda2022climate,cadiou2023challenges,faranda2023persistent}.  
The above approaches can be complemented by narratives that describe the physical mechanisms that led to an extreme weather event, namely storylines~\cite{hazeleger2015tales,shepherd2016common,lloyd2021climate}. Storylines support a detailed understanding of the factors that contributed to an event, and how they may have been influenced by human-induced climate change. For example, a storyline for a heatwave event might describe how a persistent high-pressure system led to hot and dry conditions, which were then exacerbated by human-induced climate change.

By combining analogues approaches and storylines, researchers can provide a complete picture of the link between climate change and extreme weather events~\cite{yiou2023ensembles}. Analogues approaches can quantify the probability of the event occurring, while storylines contextualise the physical mechanisms behind the event. This can help policymakers and the public to better understand the risks of climate change, and to develop strategies for adapting to and mitigating these risks.

\subsection{Mathematics for non-autonomous systems}

    Despite the success of dynamical systems  attribution-based techniques, the mathematical justification of these approaches is still largely missing (see Sect. \ref{sect:limitations}). For several physical processes it is unclear whether there exists some invariant or stationary measure which could be used for computing statistics of extreme or rare events. In this context it can be useful to return to discrete dynamical systems, for which recent results pave the way for a rigorous application of EVT-based dynamical systems approaches in an non-stationary setting. We specifically consider the following two non-stationary systems: {\em sequential} and {\em non-autonomous}. They are defined by concatenating maps chosen in some set, usually in the close neighborhood of a given map. As a probability measure one typically takes some ambient measure like Lebesgue. \\

The extreme value theory and the extremal index must then be redefined. Sect. 4.5 of \cite{freitas2017extreme} presents an example of a sequential system modeled on maps chosen in the neighborhood $\mathcal{U}$ of a given
$\beta$-transformation, say $T_{\beta_0}.$ According to suitable choices of $\mathcal{U}$ it is possible to show either that the
EI of the unperturbed map and of the sequential system are the same, or the two differ, in particular when the elements of the concatenation are far enough from $T_{\beta_0}.$ In this case the EI is simply $1$. A second relevant example is in the Appendix of \cite{caby2020extreme}. There, sequentially composed maps of the one-dimensional torus of the type $T_i=2x+b_i$-mod $1,$ are considered. The constants $b_i$ and the Dirac masses $p_i$ associated to each $T_i$ vary randomly respectively in some spaces $B_k,  P_k, k=1,2...$ which in turn change every $10$ temporal steps. The empirical distribution of the number of visits in a ball centered at the origin is in perfect agreement with a
P\`olya-Aeppli distribution.\\

The second set of systems we consider are non-autonomous random (quenched) dynamical systems. They are constructed by taking a deterministic driving map $\sigma$ on the space $E$ and preserving a probability measure $\mathbb{P}$
which codes a family of transformations
$T_{\omega}$ for $\omega \in E$ on the space $M$ via the composition rule
$ T_{\omega}^n(x)=T_{\sigma^{n-1}}\circ\cdots\circ T_{\omega}(x).$ One can prove the existence of sample measure $\mu_{\omega},$ verifying, for each measurable set $A\in M: \mu_{\omega}(T^{-1}_{\omega}(A))=\mu_{\sigma \omega}(A).$ These measures  describe the statistical properties in $M$ and they do not give rise to stationary processes. The first applications of EVT to these random systems are given in \cite{freitas2017extreme,freitas2020point}. A recent article \cite{atnip2022perturbation} further developed a new spectral approach for a quenched extreme value theory
that considers random dynamics with general ergodic invertible driving $\sigma$, and random observations. It also provided a general formula for the computation of the extremal index which will be also random. To give a flavour of such results, we quote here the following example, taken from \cite{atnip2022perturbation}. Consider the bi-infinite sequence $\omega=\{\dots, \omega_{-1}, \omega_0, \omega_1, \dots\},$ where each $\omega_l$ takes values in the finite alphabet $\{1,\dots, d\},$  and move it with the shift $\sigma.$ Each symbol will be taken with the same probability $1/d.$  We then take $d$ maps $T_1,\dots, T_d$, and random  {\em rare} sets of the form $B_{\omega,n}:=B(v(\omega_0), e^{-z_n(\omega)}).$ Notice that the centers of these balls, $v(\omega_0),$ are random ($\omega_0$ denotes the $0$-th coordinate of $\omega$),   and we also allow random radii  satisfying  the scaling $2e^{-z_n(\omega)}=\frac{t+\xi_{\omega,n}}{n},$ $t$ being a non-random constant and $|\xi_{\omega,n}|$ bounded uniformly in $n$ and $\omega$ and going to zero when $n\rightarrow \infty$ and for $\mathbb{P}$-a.a. $\omega.$ We then define the first random hitting time as
$$
\tau_{\omega,n}(x)=\inf\{k\ge 1, T^k_{\omega}(x)\in B_{\sigma^k\omega,n}\},
$$
which gives the first time the random orbit of $x$ enters an element of the sequence of random balls $B_{\sigma^k\omega,n}.$ This covers, for instance, the interesting case of rare sets which are known with a limited precision, or cases when the iterations at each step are affected by some disturbance.
For particular choices of the maps $T$ for which all the sample measures coincide and are equal to $\eta,$ we have
$$
\lim_{n\rightarrow \infty} \eta\{\tau_{\omega,n}>n\}=e^{-t\int \theta_{\omega}d\mathbb{P}(\omega)},
$$
where $\theta_{\omega}$ is the random EI, defined by suitable random generalizations of formulae (\ref{EI}) and (\ref{QQ}), see \cite{atnip2022perturbation} for the details. It is interesting to note that the previous example can be worked  out  in such a way that the expectation of $\theta_{\omega}$ is strictly less than $1,$ thus showing that we could have the formation of clusters even in the absence of periodic or invariant  structures, which could not persist forever in the presence of noise.\\

The above idealised examples show that the behaviour of non-stationary systems can be described by EI distributions, which depend on the systems' non-stationarity characteristics and on the (stochastic) system's realisation. This underscores the existence of an extreme value law for such systems, which opens the possibility of computing other dynamical quantities beyond the EI. These notions may in the future be expanded to more complex real-world systems.\\


\section{Limitations}
\label{sect:limitations}
\begin{figure*}[t]
  \noindent\includegraphics[width=1.0\textwidth]{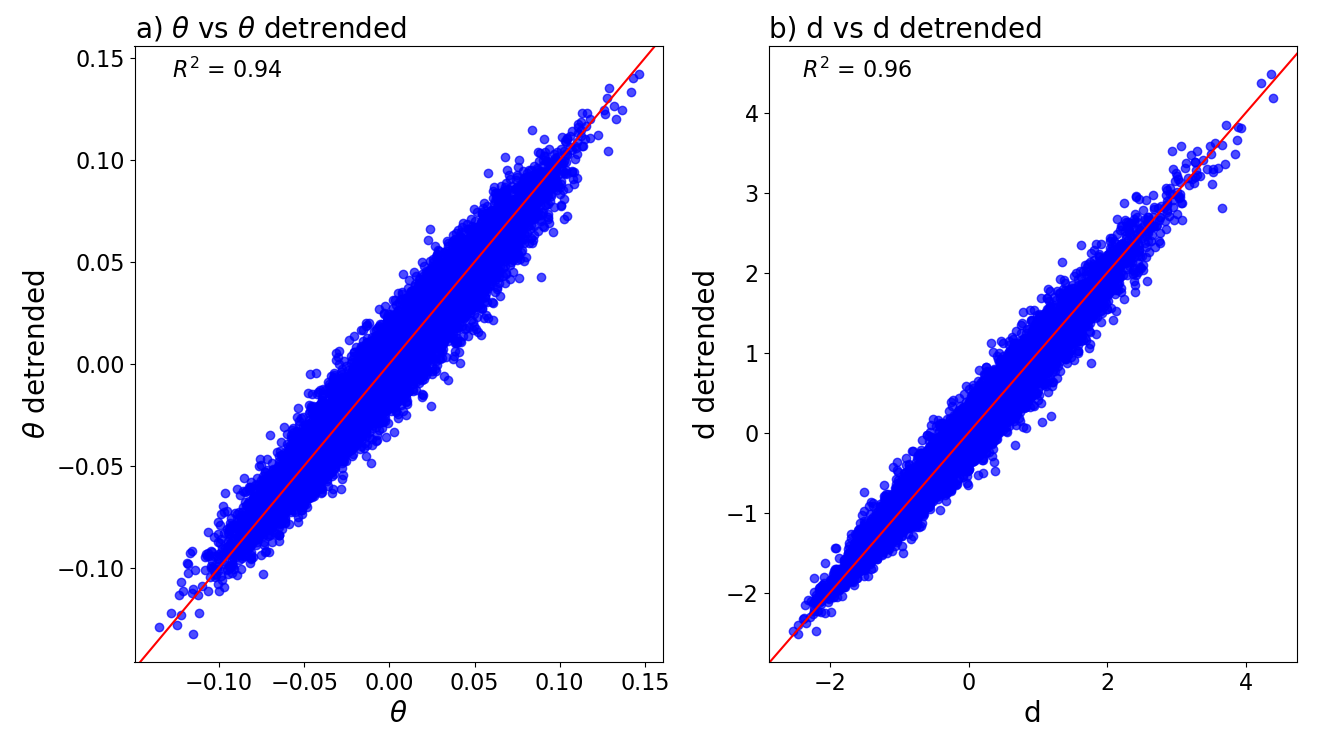}\\
  \caption{\textbf{Local dimension and persistence applied to trended and detrended climate data}. Local dimension $d$ and persistence $\theta$ computed on ERA5 Z500 data over 25–75$^\circ$N, 15$^\circ$W–50$^\circ$E and 1979-2018. The scatterplots show values computed on the raw Z500 field and on a Z500 field where a linear trend has been removed from every gridpoint. The plots show anomalies computed relative to daily climatology smoothed with a 31-day running mean.}
\label{fig:nonsta}
\end{figure*}

It is crucial to acknowledge the limitations inherent to the methodologies presented in this paper. Indeed, as already partly touched upon in Sect. II, many of the theoretical results for sampling properties of dynamical systems rely on assumption that are seldom respected by geophysical data, such as stationarity and hyperbolicity. We however note that non-stationary and/or non-hyperbolic systems may sometimes be modelled by leveraging stochastic approaches, as discussed in Sect. III. Stationarity is fundamental for any application of extreme value theory (EVT). This poses a challenge in forced systems, such as the climate system or weather patterns. With the advent of global warming, these systems have become non-stationary, including in the statistics of extreme events, questioning the validity of a stationary distribution as asymptotic limit. In practice, one can incorporate trends in the GEV parameters (mainly location and scale)~\cite{beirlant2012overview}. Complementary to this approach, many of the metrics introduced in this perspective are capable of highlighting the non-stationary behavior of long trajectories of geophysical systems \citep{faranda2019hammam,faranda2023atmospheric}. Moreover, empirical tests on non-stationary climate data emphasise the very strong correlation between metrics computed on raw versus detrended data. We present in Fig. \ref{fig:nonsta} an example for the local dimension $d$ and persistence $\theta$ computed on a raw and detrended Z500 dataset, the latter being a climate variable with a very marked trend (see e.g. \citep{lee2024dynamical}, who find a decadal trend of almost 6m area-averaged over a large mid-latitude domain). Both indicators show an extremely high correlation, sufficient to limit qualitative differences in the conclusions that one would draw for typical climate science applications. Indeed, using this same data \citep{holmberg2023link} recently showed that atmospheric analogues were not markedly affected by non-stationarity. This suggests that the techniques presented in this review can in many cases be applied to weakly non-stationary cases, namely whenever the system does not drift too rapidly away from the original attractor. This conclusion is also supported by the results of \cite{buschow2018local}, who applied the local dimension to a stationary climate setting and concluded that the results they found were "broadly consistent with the results for reanalysis data sets" (i.e. a non-stationary setting) presented by other authors. This empirical evidence, which we recognise is yet to be grounded in corresponding theoretical results, opens avenues for further mathematical research in this field.

A second assumption of the approaches that we present for sampling properties of dynamical systems is hyperbolicity. Geophysical systems often exhibit characteristics like intermittency, long-range correlations and multiple timescales, which diverge from the assumption of hyperbolicity \citep{ghil2020physics}. However, recent research has highlighted how many of the approaches we present in this study can inform on non-hyperbolic behavior \citep{faranda2019attractor,caby2019generalized,alberti2022novel}. Departures from hyperbolicity give rise to a rich phenomenology, characterized by nonlinear interactions and emergent behavior, in many cases associated with geophysical extreme events \citep{lucarini2020new}. While we recognise that there is no formal proof that the tools we discuss are applicable to non-hyperbolic systems, empirical attempts show that they can provide information on such systems. We thus argue for the need to investigate their applicability in the grey area beyond the asymptotic limits. An analogy may be drawn to the notion of penultimate distributions in EVT, which are used to formally account for finite-length timeseries \cite{gomes1999approximation}. Thus, we argue that the implications of deviations from hyperbolicity should be studied further rather than being viewed as a priori precluding the applicability of the methodologies we discuss here. 

We note several other challenges in applying EVT, large deviations, recurrence theory for dynamical systems and related approaches to geophysical extremes. As noted by~\cite{datseris2023estimating}, the application of EVT to dynamical systems is problematic to falsify. Without the ability to test hypotheses against contradictory evidence, long-standing controversies such as that surrounding the fractal dimension of a "global climate attractor" persist unresolved~\cite{lorenz1991dimension}. A further challenge is that EVT typically assumes independence between extreme events, which may require intermediate declustering steps. However, this leads to actively suppressing correlations that may issue from processes of physical interest. Moreover, ignoring these dependencies may lead to biased estimations of extreme event probabilities for example when looking at future climate projections. Similarly, most geophysical systems are high-dimensional and can only be described by considering a large number of variables simultaneously. The work we highlight in this study includes some multivariate approaches (e.g. Sect. \ref{sect:Compound}), but these are mostly applicable to small numbers of variables. EVT also relies on the availability of sufficient data to accurately estimate extreme event probabilities, but in many cases geophysical data, especially for rare or unprecedented events, may be sparse. Additionally, extreme events in geophysical systems often exhibit nonlinear dynamics and can occur across a wide range of temporal and spatial scales, requiring sophisticated modeling techniques to describe them. However, these models may contain uncertainties or inaccuracies~\cite{kuzzay2017new,faranda2017stochastic} , which can propagate to the application of the methodologies we present in this paper. Lastly, EVT typically requires the selection of a threshold above which events are considered extreme, and choosing an appropriate threshold can be challenging, influencing the results obtained from the analysis~\cite{passow2019rigorous}. Some of these limitations are not specific to geophysical extremes, and addressing them relies on broad advances in several disciplines such as statistics, mathematics and physics. We provide a flavour of some potential future research directions in Sect. \ref{sect:perspectives}.

\section{Perspectives}
\label{sect:perspectives}

Geophysical extremes, spanning phenomena as diverse as thunderstorms, tropical cyclones, earthquakes, and geomagnetic storms, encompass a broad array of spatial and temporal scales. Thunderstorms exhibit fine-scale microphysical processes occurring at the micrometer level, juxtaposed with the kilometer-scale dynamics of towering storm clouds or the scales of up to thousands of kilometers of tropical cylones. The pursuit of understanding geophysical extremes thus exemplifies the mathematical intricacies and dynamical complexities inherent to multiscale dynamics. To elucidate the multiscale fabric of these events, mathematical models must bridge multiple magnitudes in both spatial and temporal scales. We envision developments in mathematical methods such as coupled dynamical systems and network theory as instrumental to capture  complex inter-scale and inter-variable dependencies. Multiscale decomposition techniques may also play a key role in further research on the topic. These involve breaking down complex geophysical systems into multiple scales, each with its own set of governing equations and dynamics, supporting an improved understanding of how small-scale processes interact with large-scale phenomena. For example, in studying thunderstorms, multiscale decomposition can reveal how microphysical processes related to nucleation within clouds interact with the broader atmospheric circulation, or how phenomena such as tornadoes and hail form in the convective clouds. This type of knowledge is essential for improving the accuracy of climate projections and weather forecasts, and more broadly for the understanding  of the physical mechanisms driving multiple geophysical extremes.

The mathematical modeling of multiscale geophysical dynamics is rooted in partial differential equations (PDEs). Nevertheless, the severe numerical constraints posed by the direct simulations of partial differential equations necessitate multiscale modeling approaches. As we have shown in this perspective, stochastic parameterizations are a useful tool to this effect, and are able to reconcile the interactions between disparate geophysical scales. Stochastic systems further provide a probabilistic framework for modeling the inherent randomness and uncertainty in geophysical extremes, making them valuable for risk assessment and prediction. Incorporating these mathematical and analytical approaches into large deviation theory, multivariate extreme value analysis, analogue-based methods and multiscale decomposition techniques will further enhance our ability to quantify, predict, and manage geophysical extreme events, ultimately improving our resilience to their impacts. We specifically view developments in multivariate extreme value analysis as crucial to capture the joint behavior of multiple geophysical variables. In the context of geophysical extremes, this would advance the study of compound geophysical extreme events, such as the co-occurrence of high wind speeds and heavy precipitation during a (tropical) cyclone, enable devising novel bias correction techniques, or improved downscaling of extreme weather events.

In addition to the mathematical challenges mentioned above, the study of geophysical extremes presents several other dimensions. One such aspect is the need to develop methods that can accurately separate the natural variability of the climate system from the effects of human-induced climate change in engendering extreme geophysical events -- namely the so-called \textit{extreme event attribution} (Sect. V.C). This requires statistical techniques that can distinguish between short-term, internal fluctuations of geophysical systems and long-term, forced trends. The above-mentioned mathematical advances for non-stationary systems can support this effort (Sect. V.D).

Another important  challenge is the need to account for the spatial and temporal dependence of extreme events. Extreme weather events often occur in clusters or specific spatial patterns, and their occurrence in one location can be associated with the occurrence of extremes in other regions or influenced by remote drivers (see Sect. \ref{sect:Compound}). Understanding the origin of spatially co-occurring extremes requires the use of advanced spatial and temporal statistical models that can account for complex dependencies within the climate system and other geophysical systems. As we have shown when discussing co-recurrence statistics (Sect.~\ref{sec:IIC},~\ref{sect:Compound}) understanding the dependence structure of extreme events is crucial for assessing compound risks, where multiple extreme events co-occur, potentially leading to cascading impacts. In this sense the analogues-based approach presented in Sect. ~\ref{sec:IIC} could be combined with multivariate extreme value analyses to assess the likelihood and potential impacts of extreme events. More generally, the work outlined in this perspective provides some initial steps in this direction, laying the bases for a more general mathematical treatment of spatially or temporally clustered geophysical extremes.


\section{Conclusions}

This paper has provided an overview of techniques for studying geophysical extreme events, focusing on the interplay between statistical physics, dynamical systems theory, and geophysics. By identifying the limitations of traditional statistical extreme value analysis techniques to study geophysical rare events, we  have motivated the introduction of new mathematical formalisms based on rare recurrences in high-dimensional systems. 

The application of these techniques to various geophysical phenomena, such as  temperature extremes, cyclones, thunderstorms and geomagnetic storms, has provided valuable insights into the underlying dynamics and physical drivers of these events. Additionally, the examination of diverse data, including climate reanalysis, numerical climate models, geomagnetic data and turbulence measurements has illustrated how approaches developed for low-dimensional systems can be applied to high-dimensional chaotic data. While significant insights have been gained regarding the dynamics of rare events in geophysical systems, further investigation is necessary to fully comprehend the underlying mechanisms and drivers.

Understanding and analyzing high-impact events in the Earth system involves addressing several challenges, spanning both the mathematical and physical sciences. In terms of mathematical challenges, one major concern is the non-stationarity of geophysical systems. The multifarious interactions and feedback loops within the Earth system introduce a further level of complexity. Another critical challenge lies in accounting for finite size effects, including the spatial and temporal resolution of the datasets used for analysis. Moreover, geophysical systems exhibit high dimensionality, necessitating an examination of how data resolution affects the identification and characterization of rare events. Properly addressing finite size effects is essential to avoid potential biases or artifacts in the analysis \cite{lucarini2014mathematical, haydn2020limiting}.


To overcome these challenges, interdisciplinary research efforts combining mathematics, physics, and geophysics are crucial. Advances in mathematical techniques, continuing to build upon the novel indicators, approaches and modeling frameworks presented in this perspective, will provide a stepping stone for future advances. In parallel with this, a comprehensive understanding of the physics behind rare geophysical events requires a combination of observational data and theoretical insights. In this context, it is crucial to devise stochastic dynamic models, which account for the effects of environmental random fluctuations and can serve as a connection between the extreme events approach and the dynamical systems modeling. We view collaborative endeavors among researchers from various disciplines as instrumental in this respect. We specifically encourage continued collaboration between researchers in statistical physics, statistics, dynamical systems theory and geophysics, whose expertise this perspective has sought to gather. In the longer term, we envision the integration of tailored machine learning algorithms and advanced data assimilation techniques with dynamical systems and statistical physics approaches to further enhance our understanding of geophysical extremes and improve their prediction capabilities. Ultimately, our hope is that these collective efforts of researchers across various disciplines will contribute to building a more resilient and sustainable future.

\section*{Acknowledgements}
This perspective has issued from discussions held during the UNDERPIN symposium held in Erice, Italy, with the support of MITI-CNRS and the Ettore Maiorana Foundation and Centre for Scientific Culture. All authors wamrly acknowledge two anonymous reviewers and the handling editor, Valerio Lucarini, for very useful comments. SV thanks the Mathematical Research Institute MATRIX, the Sydney Mathematical Research Institute (SMRI), the University of New South Wales, and the University of Queensland for their support and hospitality. DF, PY, RN and EC received support from the European Union’s Horizon 2020 research and innovation programme under grant agreement No. 101003469 (XAIDA). DF, PY, RN and GM received further support from  the European Union’s Horizon 2020 Marie Sklodowska-Curie grant agreement No. 956396 (EDIPI) and the LEFE-MANU-INSU-CNRS grant "CROIRE". The research of LC was supported by the European Union’s Horizon 2020 Marie Sklodowska-Curie grant agreement No. 101065985 (CYCLOPS). We also acknowledge useful discussions with the MedCyclones COST  Action  (CA19109) community.  The research of SV was supported by the project {\em Dynamics and Information Research Institute} within the agreement between UniCredit Bank and Scuola Normale Superiore di Pisa and by the Laboratoire International Associ\'e LIA LYSM, of the French CNRS and  INdAM (Italy).  SV was also supported by the project MATHAmSud TOMCAT 22-Math-10, N. 49958WH, of the French  CNRS and MEAE. PY and DF received support from the grant ANR-20-CE01-0008-01 (SAMPRACE)\\

\section{Appendix}
We collect here some formulas and quantities introduced in section II.B. First of all  the quantities $\lambda_l$ given in (\ref{eq:EI}) are formally defined as
\begin{equation}
\lambda_l:=\lim_{\Delta\rightarrow \infty}\lim_{n\rightarrow \infty}\frac{\mu(x; \sum_{j=0}^{\Delta}{\bf 1}_{U_n}( T^j(x))=l)}{\mu(x; \sum_{j=0}^{\Delta}{\bf 1}_{U_n}(T^j(x))\ge 1)},
\end{equation}
provided the limits exist.\\

We now quote the analytic expression of the quantities $q_{k,n}$ entering the definition of the extremal index (\ref{EI}):
\begin{equation}\label{QQ}
q_{k,n}=\frac{\mu(x\in U_n, T(x)\in U_n^c,\cdots, T^k(x)\in U_n^c, X_{k+1}\in U_n)}{\mu(U_n)}.
\end{equation}

We finally quote the mass distribution of the P\`olya-Aeppli distribution

\begin{equation}
\tilde{\nu}(\{k\})=
e^{-(1-p)t}\sum_{j=1}^k\binom{k-1}{j-1}\frac{((1-p)^2t)^j}{j!}p^{k-j}.
\end{equation}

\bibliography{bib_rare_events}

\end{document}